\documentclass[11pt]{article}

\usepackage[final]{acl}

\usepackage{times}
\usepackage{latexsym}
\usepackage[T1]{fontenc}
\usepackage[utf8]{inputenc}
\usepackage{microtype}
\usepackage{inconsolata}
\usepackage{graphicx}
\usepackage{amsmath} 
\usepackage{booktabs} 
\usepackage{multirow}
\usepackage{enumitem}           
\usepackage{graphicx}
\usepackage{xcolor}
\usepackage{amssymb}
\usepackage{xspace}
\usepackage{xcolor}
\usepackage{listings}
\usepackage{fancyhdr}
\usepackage[most]{tcolorbox}
\usepackage[svgnames]{xcolor}
\usepackage{enumitem}

\AtBeginDocument{\hypersetup{hyperfootnotes=false}}

\NewDocumentCommand{\boyan}{ mO{} }{\textcolor{blue}{\textsuperscript{\textit{Boyan}}\textsf{\textbf{\small[#1]}}}}

\title{ROSE: An Intent-Centered Evaluation Metric for NL2SQL}

\author{
  Wenqi Pei\textsuperscript{1,2,}\thanks{Equal Contribution.},
  Shizheng Hou\textsuperscript{2,}\footnotemark[1],
  Boyan Li\textsuperscript{1,}\footnotemark[1],
  Han Chen\textsuperscript{2},
  Zhichao Shi\textsuperscript{1},
  Yuyu Luo\textsuperscript{1,}\thanks{Corresponding author.}\\
  \\\ 
  \textsuperscript{1} The Hong Kong University of Science and Technology (Guangzhou) \\ 
  \textsuperscript{2} National University of Singapore 
}

\begin{document}

\newcommand{\metric}{ROSE\xspace}
\newcommand{\ablation}{ROSE w/o Refuter\xspace}
\newcommand{\dataset}{ROSE-VEC\xspace}

\begin{NoHyper}
\maketitle
\end{NoHyper}

\begin{abstract}
Execution Accuracy (EX), the widely used metric for evaluating the effectiveness of Natural Language to SQL (NL2SQL) solutions, is becoming increasingly unreliable. It is sensitive to syntactic variation, ignores that questions may admit multiple interpretations, and is easily misled by erroneous ground-truth SQL. To address this, we introduce \textbf{\metric}, an intent-centered metric that focuses on whether the predicted SQL answers the question, rather than consistency with the ground-truth SQL under the reference-dependent paradigm. ROSE employs an adversarial Prover-Refuter cascade: SQL Prover assesses the semantic correctness of a predicted SQL against the user's intent independently, while Adversarial Refuter uses the ground-truth SQL as evidence to challenge and refine this judgment. On our expert-aligned validation set \textbf{\dataset}, \metric achieves the best agreement with human experts, outperforming the next-best metric by nearly 24\% in Cohen's Kappa. We also conduct a large-scale re-evaluation of 19 NL2SQL methods, revealing four valuable insights. We release \metric and \dataset to facilitate more reliable NL2SQL research\footnote{https://github.com/CedricPei/ROSE}.
\end{abstract}

\section{Introduction}
\label{sec:intro}

Natural Language to SQL (NL2SQL) translates user questions into executable SQL queries over a given database~\cite{DBLP:journals/pvldb/LuoLFCT25,DBLP:journals/tkde/LiuSLMJZFLTL25, zhu2026surveydataagentsemerging} and supports a range of data analysis tasks~\cite{zhang2021tadoc,deepeye_icde,deepeye_sigmod}. Recent progress in NL2SQL has been remarkable, driven by increasingly capable large language models (LLMs) ~\cite{deepeye_sql, DBLP:journals/corr/abs-2503-22402,DBLP:journals/corr/abs-2505-04671,pei-etal-2025-feather}. However, this rapid growth has exposed a key bottleneck: existing evaluation metrics struggle to capture whether the predicted SQL is semantically correct with respect to the user's intent ~\cite{Li2024SuperSQL,hou2026nl2sqlbench}.

The core metric in the field, Execution Accuracy (EX), deems a predicted SQL correct only when its execution result matches a single ground-truth SQL. This surface-level check causes EX to be unreliable in three recurring cases: (i) Realization variance when the underlying logic is identical (e.g., projection order, value formatting). EX cannot accommodate such predictions across different output representations. Structural audits report false negatives up to 28.9\% caused by non-canonical but correct forms~\cite{ascoli2024etm}; (ii) Multiple valid interpretations for ambiguous questions. EX misses these reasonable predictions that deviate from the ground-truth SQL but still correctly answer the question. Ambrosia~\cite{saparina2024ambrosia} reports that more than half of failures arise from ambiguity, and Sphinteract \cite{zhao2025sphinteract} further underscores the severity of the problem; (iii) Erroneous ground-truth SQL in large-scale benchmarks. EX propagates these annotation errors to all evaluated predictions. NL2SQL-BUGs \cite{liu2025nl2sqlbugs} reports a 6.91\% ground-truth SQL error rate on BIRD Dev. In our audit, approximately 25\% of sampled items were flagged as wrong by at least one annotator. These issues indicate that EX favors matching a particular reference SQL rather than achieving the intent behind the question.

Several improved metrics have been proposed. Structure-aware measures \cite{ascoli2024etm, zhan-etal-2025-towards} allow for syntax-level variants via complex normalization, which reduces superficial mismatches, but still judges proximity to a reference rather than whether the prediction answers the question's semantics. LLM-based judges \cite{kim2025flex, llmsqlsolver} improve alignment with expert assessments, but most protocols remain reference-dependent, checking consistency primarily against a single ground-truth SQL. They are insufficiently tolerant of legitimate ambiguity and are easily misled by the flawed ground-truth SQL. In light of these limitations, a paradigm shift toward less reference-dependent evaluation is becoming advisable.

Therefore, we introduce \textbf{\metric} (\textbf{R}eas\textbf{O}ning \textbf{S}cor\textbf{E}). It is a new metric designed to measure the consistency between the underlying intent of the question and the reasoning process embodied by the prediction. This intent-centered evaluation is realized via a Prover-Refuter cascade powered by reasoning models. The \textit{SQL Prover} makes an independent judgment using only the question and the database information, without accessing the ground-truth SQL. The \textit{Adversarial Refuter}, with access to the ground-truth SQL, treats it as evidence to challenge the Prover's acceptance, contrasting it with the prediction to expose decisive mismatches. Through this analysis, it can also tag cases as ambiguous questions or ground-truth errors. This cascade reduces reference anchoring while tempering over-permissiveness and exposes dataset issues. 

We construct \textbf{\dataset}, a \textbf{V}alidation dataset with \textbf{E}xpert \textbf{C}onsensus to assess metric validity. On this dataset, ROSE outperforms the closest competitor by 24\% in agreement and 14\% in accuracy, demonstrating better alignment with user intent.

Building on \metric, we perform a large-scale re-evaluation of 19 NL2SQL methods. Our analysis yields four key insights: (i) Base model capability, not system-level engineering, is the primary performance driver; (ii) As models advance, a widening gap between semantic correctness and reference matching signals an evaluation crisis; (iii) This divergence largely stems from benchmark flaws, namely ground-truth errors and question ambiguities; and (iv) Fine-tuning narrows this gap by aligning models to a dataset's stylistic conventions.

We summarize our contributions as follows:
\vspace{-2mm}
\begin{itemize}[leftmargin=*, itemsep=0pt]
\item We introduce \metric, an intent-centered metric for NL2SQL evaluation that leverages an adversarial Prover-Refuter cascade, achieving the best agreement with expert judgment.
\item We construct and release \dataset, a validation dataset of 585 expert-consensus samples, complete with detailed annotations, to enable rigorous validation of NL2SQL metrics.
\item We conduct a large-scale re-evaluation of 19 methods, distilling four key insights that provide important guidance for future NL2SQL research.
\end{itemize}

\section{Preliminary}
% This section establishes the formulation for the NL2SQL evaluation problem. We introduce the necessary notations and define a theoretical framework for correctness.

\subsection{Problem Formulation}
The task of NL2SQL is to translate a NL question $Q$ into SQL $S$, conditioned on a given database and its content, collectively denoted $D$. Given $Q$ and $D$, an NL2SQL method generates a predicted SQL $S_p$. For evaluation, benchmark datasets provide a corresponding ground-truth SQL $S_g$. Executing these queries against the database $D$ yields their respective result sets $E_p$ and $E_g$.

After \(S_p\) passes syntactic checks, we evaluate semantic correctness under a set of acceptance criteria \(C\) that encode user-specific rules on schema validity and alignment with \(Q\) (e.g., tolerance for duplicates or NULL). Any executable \(S_p\) that violates \(C\) is considered incorrect.

\subsection{Ideal Evaluation}
The ideal evaluation of a predicted SQL $S_p$ assesses whether it correctly captures the user's intent. We formalize this with a theoretical judgment function, $I$, which decomposes the problem into two conditions: syntactic validity and semantic correctness.
$$I(Q, S_p, D, C) =\sigma_{\text{syn}}(S_p \ |D) \; \land \; \sigma_{\text{sem}}(S_p, Q \ | C)$$

Here, $\sigma_{\text{syn}}$ verifies the query's syntactic validity against the database $D$, while $\sigma_{\text{sem}}$ assesses if the query's logic semantically captures the intent of the question $Q$ according to the criteria $C$.

While $\sigma_{\text{syn}}$ is trivial to implement, a perfect $\sigma_{\text{sem}}$ is computationally infeasible \cite{abiteboul1995foundations, he2024verieql}. Practical methods approximate this by measuring the similarity of $S_p$ and $S_g$ through structural or execution result comparison.

\section{Related Work}
The evaluation of NL2SQL has evolved significantly. We categorize the landscape of evaluation metrics into two primary types: deterministic metrics and LLM-based metrics. The formal mathematical definitions for the metrics are available in the Appendix \ref{sec:math-nl2sql-metrics}.

\subsection{Deterministic Metrics}
Deterministic metrics evaluate query correctness using predefined, rule-based algorithms, forming the foundation of standard NL2SQL benchmarks.

\textbf{Exact Match (EM)} \cite{yu2018spider}, also known as String Match (SM), is the strictest metric, requiring $S_p$ to be character-for-character identical to $S_g$ after normalization. To provide partial credit, \textbf{Component Match (CM)} \cite{yu2018spider} evaluates correctness at the clause level. It scores $S_p$ by calculating the proportion of its components that correctly match $S_g$.

\textbf{Execution Accuracy (EX)} \cite{yu2018spider} verifies the equivalence of execution results. $S_p$ is deemed correct if it produces the same result as $S_g$. \textbf{Enhanced Tree Match (ETM)} \cite{ascoli2024etm} operates on a structural level. It parses $S_p$ and $S_g$ into Abstract Syntax Trees (ASTs). A match is declared when their normalized forms are structurally equivalent. Other approaches include \textbf{Exact Set Matching (ESM)} \cite{yu2018spider}, which compares unordered sets of keywords and arguments, and \textbf{Distilled Test Suite} \cite{zhong2020semanticevaluationtexttosqldistilled} designed to probe specific SQL capabilities.

% Its primary drawback is brittleness to minor, semantically inconsequential syntactic variations.

% However, EX is susceptible to false positives on sparse databases and false negatives from non-semantic differences in output formatting.

% ETM mitigates false negatives of EM while avoiding false positives of EX.

The reliance of deterministic metrics on a single and sometimes flawed $S_g$ penalizes valid alternatives and misrepresents method performance, thus necessitating more flexible evaluation metrics.

\vspace{-1.6mm}
\subsection{LLM-based Metrics}
\label{sec:llm-based-metrics}
A new evaluation type has emerged that employs Large Language Models (LLMs) as semantic judges to assess query correctness. Unlike deterministic metrics, this approach aims to leverage the reasoning capabilities of LLMs to provide a more realistic measure of utility.

\textbf{LLM-SQL-Solver}~\cite{llmsqlsolver} directly prompts an LLM to determine whether a predicted SQL is equivalent to $S_g$. It employs prompting strategies such as Miniature \& Mull to find counterexamples and Explain \& Compare to analyze logic. To reduce false positives and negatives of EX, \textbf{FLEX}~\cite{kim2025flex} leverages meticulously crafted prompts that provide complete context, guiding an LLM to deliver adequacy judgments. Frameworks from industry such as Defog.ai and Arize also use LLMs for evaluation \cite{defog2023sqleval_blog, arize2024_text_to_sql_llm_judge}.

However, existing LLM-based metrics still rely on $S_g$, which can be misleading and unfairly penalize valid alternatives. Our work addresses this by transforming the role of $S_g$ from a standard reference into an adversarial challenge.

% Although these LLM-based approaches offer greater flexibility, many still inherit the single reference issue by using $S_g$ as the primary reference for the LLM to judge against. This means that the evaluation can still be misled by a flawed or non-unique ground-truth SQL, anchoring the otherwise powerful reasoning of the LLM to a potentially unreliable benchmark.

% Therefore, the academic community urgently needs an evaluation metric that can leverage the powerful reasoning capabilities of LLMs without inheriting the fragility of a single, unreliable reference. Our work addresses this critical gap for the first time by fundamentally reshaping the role of the gold standard in the evaluation process—from a 'reference' to an 'adversarial challenge.'

\section{Methodology}
% Our proposed metric, \metric, redefines the NL2SQL evaluation by treating $S_g$ not as a gold standard, but as adversarial evidence to challenge $S_p$. 
% This approach addresses the brittleness of metrics like EX against valid query variations or erroneous ground-truth SQL. 
% After ensuring a query is syntactically correct, \metric employs a two-stage process: a ground-truth independent SQL Prover for an initial sufficiency check, and an Adversarial Refuter that uses $S_g$ to refute the Prover's judgment.

\subsection{SQL Prover}
To overcome the limitations of reference-dependent metrics, a mechanism is required to validate the reasoning of a predicted SQL against the user's intent without reliance on $S_g$. To this end, we introduce SQL Prover, designed to assess the semantic correctness of a query independently. It is invoked only when $S_p$ is syntactically valid and $E_p$ differs from $E_g$. It evaluates whether $S_p$ satisfies the user's intent expressed in $Q$ under the acceptance criteria $C$. Formally, SQL Prover function $P$ outputs a boolean judgment $j_p$ and a rationale $R$:
$$Pro(Q, S_p, E_p \ |D, C) \rightarrow (j_{p}, R)$$

Detailed acceptance criteria and instructions are provided in Appendix~\ref{sec:prover-prompt}.

\subsection{Adversarial Refuter}
However, SQL Prover's complete independence from the ground-truth risks being overly permissive, failing to leverage signals (albeit noisy) within $S_g$. Hence, we introduce Adversarial Refuter. It uses $S_g$ not as a reference to match, but as a source of evidence to challenge and potentially refute SQL Prover's affirmative judgment on $S_p$.

\subsubsection{$E_p \equiv E_g$: Suppressing False Positives}
When execution results match, SQL Prover is bypassed, and the Refuter acts as a critical safeguard. By directly comparing the reasoning of $S_p$ and $S_g$, it identifies cases of coincidental correctness. This prevents false positives from a flawed prediction and, just as importantly, labeling erroneous $S_g$.

\subsubsection{$E_p \not\equiv E_g$: Challenging SQL Prover}
In cases where SQL Prover approves $S_p$ despite $E_p$ conflicting with $E_g$, Adversarial Refuter arbitrates the conflict. It pinpoints the semantic divergence between the reasoning of $S_p$ and $S_g$. It then re-evaluates this divergence against the user's intent in $Q$ to determine which logic is more faithful. This arbitration may refute SQL Prover's approval, flag $S_g$ as erroneous, or accept both queries as valid interpretations of an ambiguous question $Q$.

Adversarial Refuter's function $Ref$ produces a boolean judgment $j_r$ (overturn for true, uphold for false) on $S_P$ and a diagnostic label $L_q$ for $Q$. The function is defined as follows:
$$Ref(Q, S_p, S_g; E_p^*, E_g^*, R^* \ | D, C) \rightarrow (j_{r}, L_q)$$
The conditional arguments ($E_p^*, E_g^*, R^*$) are invoked only when $E_p \not\equiv E_g$. 

\begin{figure}[htbp]
\centering
\includegraphics[width=\linewidth]{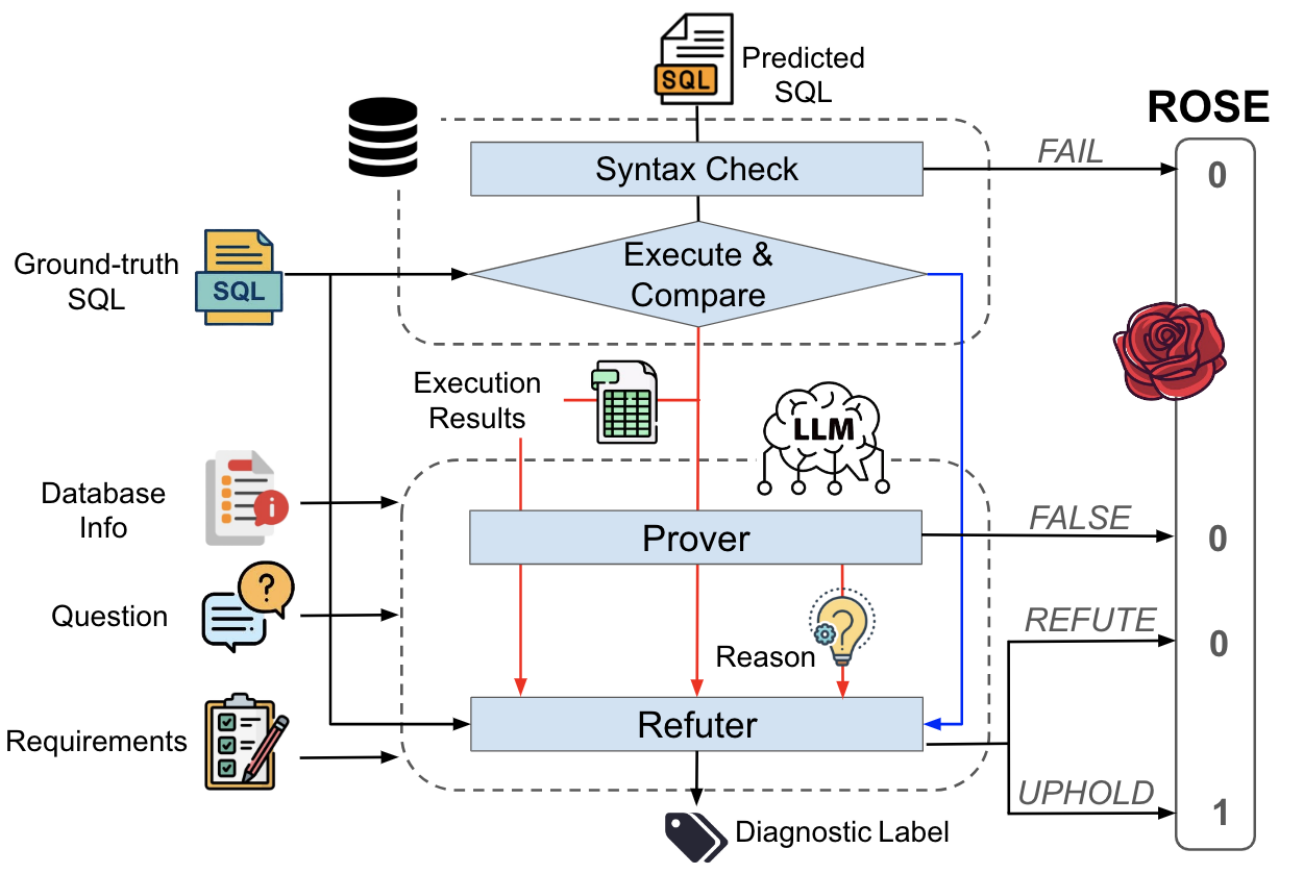}
\caption{ROSE Scoring Cascade. The \textcolor{red}{red} path indicates the workflow when execution results do not match, while the \textcolor{blue}{blue} path is followed when they do.}
\label{fig:rose_framework}
\vspace{-3mm}
\end{figure}

\subsection{\metric}

% The integrated \textbf{\metric} workflow first filters for executable queries. If execution results of $S_p$ and $S_g$ match, $S_p$ is sent directly to the Refuter. If they differ, $S_p$ first goes to the Prover; a "pass" advances it to the Refuter, while a "fail" terminates the evaluation. A query is deemed correct only if it passes all applicable stages. 

\metric is determined by the workflow illustrated in Figure~\ref{fig:rose_framework}. A predicted SQL must first be executable. If its execution result differs from the ground-truth SQL, it must pass SQL Prover's independent evaluation. Finally, it must withstand the adversarial challenge from Adversarial Refuter, which uses the ground-truth as counter-evidence. Failure at any of these stages results in a score of 0. Only a query that successfully navigates this entire cascade earns a final score of 1. The mathematical definition for ROSE is in Appendix~\ref{sec:math-rose} and its comparison with other NL2SQL metrics is in Appendix~\ref{sec:appendix_comparison}.

\section{Experiments: Validating \metric}

\subsection{Setup}

\subsubsection{\dataset}
To enable rigorous validation of NL2SQL metrics against expert judgments, we construct a human-labeled dataset \dataset consisting of 585 NL-SQL pairs. It includes 263 pairs from outputs of DAIL-SQL~\cite{Gao2024DAILSQL}, RSL-SQL~\cite{cao2024rslsql}, and Super-SQL~\cite{Li2024SuperSQL} on Spider Test~\cite{yu2018spider} and 322 pairs from outputs of TA-SQL~\cite{qu2024tasql}, CSC-SQL~\cite{sheng2025cscsql}, OpenSearch-SQL~\cite{xie2025opensearchsql}, Alpha-SQL~\cite{li2025alphasql}, and RSL-SQL on BIRD Dev. Each output is independently judged by two of the five experts using our evaluation interface, and we retain only cases with exact agreement. The resulting dataset stores instances $(Q, D, S_g, S_p, Y)$, where $Y$ denotes the consensus semantic correctness of $S_p$ for $Q$. Details of the annotators are in Appendix~\ref{sec:annotator-details}, and the evaluation interface is shown in Appendix~\ref{sec:eval-interface}. \dataset and the interface constitute valuable resources for follow-up studies.

\subsubsection{Validation Metrics}
We evaluate agreement between NL2SQL metrics and expert labels, treating ``correct'' as the positive class. We report four validation metrics, with formulations deferred to Appendix~\ref{sec:appendix_metrics}.

\vspace{1mm}

\noindent\textbf{Cohen's Kappa ($\kappa$)} \cite{cohen1960kappa} is our primary metric. It measures agreement beyond chance and is robust under skewed distributions.

\vspace{1mm}

\noindent\textbf{Accuracy (ACC)} is the proportion of correctly judged instances. It is simple to interpret but can be misleading when the classes are imbalanced.

\vspace{1mm}

\noindent\textbf{Matthews Correlation Coefficient (MCC)} \cite{matthews1975mcc} is a correlation coefficient between ground-truth and predicted labels that is reliable when class sizes differ substantially.

\vspace{1mm}

\noindent\textbf{F1} \cite{vanrijsbergen1979ir} is the harmonic mean of precision and recall on the positive class, assessing how well a judge identifies valid SQL queries.

\subsubsection{NL2SQL Metrics}
We consider two categories of NL2SQL metrics. Deterministic metrics include EX, EM, and ETM. LLM-based metrics include LLM-SQL-Solver, FLEX, and our proposed ROSE. We also report an ablation baseline \ablation, which drops the Refuter and uses only the SQL Prover. More ablation studies are provided in Appendix~\ref{sec:appendix-ablations}.

\subsubsection{Reasoning Models}
Following the naming convention in Appendix~\ref{sec:naming-convention}, we instantiate three reasoning models for LLM-based metrics: closed-source OpenAI o3-2504~\cite{openai2025o3} and Google Gemini-2.5 Pro-2506~\cite{google2025gemini25pro}, and open-source DeepSeek-R1-2505~\cite{guo2025deepseekr1}. Results on more open-source backbones are in Appendix~\ref{sec:appendix-opensource}.

\subsection{Results}

\begin{table*}[!t]
\centering
\begingroup
\fontsize{10pt}{12pt}\selectfont
\setlength{\tabcolsep}{14pt}
\renewcommand{\arraystretch}{1.15}
\caption{Performance of NL2SQL metrics on \dataset. Within each block, \textbf{Bold} indicates the best metric, \underline{Underline} indicates the second best. Open-source models are represented with {\small\unlockicon{}}, while closed-source models are represented with {\small\lockicon{}}. Separate results on the BIRD and Spider splits are in Appendix~\ref{sec:appendix_metric_effectiveness}.}
\label{tab:verification-metrics}
\begin{tabular}{l l r r r r}
\toprule
\textbf{Backbone} & \textbf{Metric} & \textbf{\(\boldsymbol{\kappa}\) (\%)} & \textbf{Acc (\%)} & \textbf{MCC (\%)} & \textbf{F1 (\%)} \\
\midrule
\multirow{3}{*}{\textbf{Deterministic}}
& EM   & 0.51  & 27.86 & 5.07  & 1.86  \\
& ETM  & \underline{6.60}  & \underline{35.56} & \underline{18.47} & \underline{20.63} \\
& EX   & \textbf{25.56} & \textbf{55.90} & \textbf{37.23} & \textbf{57.00} \\
\midrule
\multirow{4}{*}{\textbf{OpenAI o3}\textsuperscript{\lockicon}}
& LLM-SQL-Solver & 12.24 & 42.05 & 25.54 & 33.92 \\
& FLEX           & 56.70 & 78.97 & \underline{62.01} & 83.31 \\
& \ablation & \underline{60.74} & \underline{85.47} & 61.46 & \underline{90.40} \\
& \textbf{ROSE}  & \textbf{80.43} & \textbf{91.79} & \textbf{81.04} & \textbf{94.16} \\
\midrule
\multirow{4}{*}{\textbf{Gemini-2.5 Pro}\textsuperscript{\lockicon}}
& LLM-SQL-Solver & 20.31 & 50.43 & 33.62 & 48.40\\
& FLEX           & 43.53 & 70.60 & 51.16 & 75.14 \\
& \ablation & \underline{48.28} & \underline{82.39} & \underline{51.68} & \underline{88.84} \\
& \textbf{ROSE}  & \textbf{69.68} & \textbf{86.84} & \textbf{71.01} & \textbf{90.41} \\
\midrule
\multirow{4}{*}{\textbf{DeepSeek-R1}\textsuperscript{\unlockicon}}
& LLM-SQL-Solver & 15.00 & 46.32 & 26.04 & 42.91 \\
& FLEX           & 29.09 & 59.32 & 39.64 & 61.86 \\
& \ablation & \underline{46.90} & \underline{79.15} & \underline{46.90} & \underline{85.75} \\
& \textbf{ROSE}  & \textbf{64.49} & \textbf{84.62} & \textbf{65.68} & \textbf{88.81} \\
\bottomrule
\end{tabular}
\endgroup
\end{table*}

\subsubsection{Metric Effectiveness}
Table~\ref{tab:verification-metrics} reveals a significant gap between deterministic metrics and expert judgment.  EX achieves only 25.56\% in Cohen's $\kappa$, confirming that it is a poor proxy for semantic correctness.

Though existing LLM-based metrics offer an improvement, they still fall short of expert-level agreement. Our ground-truth independent ablation \ablation already surpasses them. This reflects the limitations of reliance on the ground-truth SQL, as discussed in Section~\ref{sec:llm-based-metrics}.

With three different backbones, \metric consistently achieves the best performance across validation metrics. ROSE$_{\textit{o3-2504}}$ reaches 80.43\% in Cohen's $\kappa$, 24\% improvement over FLEX$_{\textit{o3-2504}}$. This demonstrates the superiority of our cascade. The substantial gain from \ablation to the full cascade suggests that adversarial use of the ground-truth is key to peak performance. We analyze representative error cases in Appendix~\ref{sec:error-analysis}.

\subsubsection{Diagnostic Effectiveness}

Beyond scoring, \metric serves as an effective diagnostic tool for identifying erroneous ground-truth SQL and ambiguous questions. To validate this capability, we manually verified the questions flagged by Refuter as having erroneous ground-truth SQL (GoldX) or being ambiguous (AmbQ) and calculated the precision of these labels.

\begin{table}[t!]
\begingroup
\fontsize{10pt}{11pt}\selectfont
\setlength{\tabcolsep}{6pt}
\renewcommand{\arraystretch}{1.1}
\caption{Precision of \metric diagnostic labelling for GoldX and AmbQ on ROSE-VEC. Separate results on the BIRD and Spider splits are in Appendix~\ref{sec:appendix_diag_precision}.}
\label{tab:verification-tags}
\begin{tabular}{l l r r}
\toprule
\textbf{Label} & \textbf{Model} & \textbf{Count} & \textbf{Precision (\%)} \\
\midrule
\multirow{3}{*}{GoldX}
& OpenAI o3        & 51  & \textbf{84.32} \\
& Gemini-2.5 Pro   & 76  & 68.42 \\
& DeepSeek-R1      & 76  & 61.84 \\
\midrule
\multirow{3}{*}{AmbQ}
& OpenAI o3        & 57  & \textbf{91.23} \\
& Gemini-2.5 Pro   & 63  & 73.02 \\
& DeepSeek-R1      & 108 & 51.85 \\
\bottomrule
\end{tabular}
\endgroup
\vspace{-2mm}
\end{table}

As detailed in Table~\ref{tab:verification-tags}, ROSE$_{\textit{o3-2504}}$ demonstrates high reliability. It achieves a precision of 84.32\% for identifying ground-truth SQL errors and 91.23\% for ambiguous questions. Although precision varies across different backbones, the strong performance of ROSE$_{\textit{o3-2504}}$ confirms that its diagnostic labels are reliable enough for automated dataset analysis and cleaning, highlighting its dual value as both a superior evaluation metric and a powerful benchmark-auditing tool.

\subsection{Efficiency and Versioning}
We optimize the time efficiency of \metric through multi-thread parallelism in Appendix~\ref{sec:appendix-latency}. To reduce monetary cost, we utilize concise prompts and route calls across stages, observing that \metric can be more cost-efficient than FLEX, as shown in Appendix~\ref{sec:appendix-cost}. In addition, to maintain long-term stability under rapidly evolving LLMs, we adopt a backbone versioning strategy with selection and update policy detailed in Appendix~\ref{sec:backbone-versioning}.

\renewcommand{\arraystretch}{1.1}          
\newcommand{\grouprow}[1]{\multicolumn{11}{c}{\small\bfseries #1}\\}  
\newcommand{\basem}[1]{\underline{#1}}    

\begin{table*}[htbp]
\centering
\caption{Benchmarking results of NL2SQL methods on BIRD mini-Dev with EX and ROSE$_{\textit{o3-2504}}$. 
\underline{Underlined} methods denote base models. The score gap is illustrated in Appendix~\ref{sec:appendix_score_gap}.}
\label{tab:bird_main}
\footnotesize
\setlength{\tabcolsep}{6pt}
\begin{tabular}{l c l | cc | cc | cc | cc}
\toprule
\multirow{2}{*}{\textbf{Method}} & \multirow{2}{*}{\textbf{Date}} & \multirow{2}{*}{\textbf{Model}} &
\multicolumn{2}{|c|}{\textbf{Simple}} &
\multicolumn{2}{|c|}{\textbf{Moderate}} &
\multicolumn{2}{|c|}{\textbf{Challenge}} &
\multicolumn{2}{|c}{\textbf{Overall}} \\
& & & EX & \metric & EX & \metric & EX & \metric & EX & \metric \\
\midrule
\grouprow{Prompting Methods}
\midrule
\basem{GPT-5}          & Aug 2025 & --             & 69.86 & 91.10 & 51.03 & 87.24 & 46.46 & 89.90 & 55.74 & 88.93 \\
Alpha-SQL-32B          & Feb 2025 & Qwen2.5-Coder  & 81.02 & 91.24 & 69.47 & 79.65 & 52.58 & 70.10 & 69.35 & 81.09 \\
OpenSearch-SQL         & Sep 2024 & DeepSeek-Chat  & 77.94 & 91.18 & 69.78 & 80.44 & 56.25 & 67.71 & 69.37 & 80.96 \\
RSL-SQL                & Oct 2024 & DeepSeek-Chat  & 74.29 & 86.43 & 59.83 & 71.79 & 52.04 & 62.24 & 62.50 & 74.15 \\
\basem{DeepSeek-Chat}  & Mar 2025 & --             & 72.60 & 84.25 & 45.93 & 69.92 & 37.62 & 60.40 & 52.13 & 72.21 \\
RSL-SQL                & Oct 2024 & GPT-4o         & 80.15 & 91.18 & 64.60 & 80.09 & 53.61 & 73.20 & 66.88 & 81.92 \\
\basem{GPT-4o}         & May 2025 & --             & 68.92 & 79.45 & 48.40 & 69.01 & 60.40 & 65.35 & 52.20 & 71.37 \\
SuperSQL               & Jul 2024 & GPT-4          & 67.65 & 72.79 & 48.66 & 69.38 & 45.83 & 48.96 & 53.73 & 61.18 \\
TA-SQL                 & May 2024 & GPT-4          & 66.67 & 71.74 & 49.15 & 54.66 & 33.67 & 44.90 & 51.06 & 57.63 \\
DAIL-SQL               & Nov 2023 & GPT-4          & 62.50 & 72.06 & 44.64 & 52.68 & 38.95 & 38.95 & 48.79 & 55.60 \\
\basem{GPT-4}          & Apr 2024 & --             & 65.75 & 76.03 & 44.03 & 67.90 & 36.63 & 49.50 & 48.98 & 66.53 \\
CoT                    & Mar 2023 & GPT-3.5        & 42.65 & 50.00 & 21.05 & 28.07 & 13.54 & 19.79 & 25.87 & 32.83 \\
C3-SQL                 & Jul 2023 & GPT-3.5        & 61.03 & 62.50 & 36.89 & 43.11 & 28.87 & 30.93 & 42.36 & 46.29 \\
\basem{GPT-3.5}        & Jan 2024 & --             & 56.16 & 68.49 & 36.21 & 44.03 & 31.68 & 40.59 & 41.22 & 50.61 \\
\midrule
\grouprow{Fine-tuned Methods}
\midrule
CSC-SQL-32B            & May 2025 & XiYan-Qwen2.5  & 85.00 & 87.14 & 71.19 & 77.54 & 53.06 & 67.35 & 71.52 & 78.27 \\
OmniSQL-32B            & Mar 2025 & Qwen2.5-Coder  & 80.00 & 86.43 & 67.23 & 78.30 & 57.14 & 72.45 & 68.92 & 79.49 \\
CodeS-15B              & Feb 2024 & StarCoder      & 64.96 & 67.15 & 46.90 & 44.69 & 39.13 & 32.61 & 50.77 & 49.01 \\
CHESS\textsubscript{(IR,SS,CG)}               & May 2024 & DeepSeek-Coder          & 70.50 & 70.50 & 58.23 & 62.87 & 45.92 & 48.98 & 59.28 & 62.24 \\
RESDSQL-3B              & Feb 2023 & T5             & 56.30 & 54.07 & 31.44 & 31.00 & 17.71 & 15.62 & 35.87 & 34.57 \\
\bottomrule
\end{tabular}
\end{table*}

% \textbf{Stability and objectivity.} To ensure stability, we fix a single unified judge for public reporting, chosen by its agreement with an expert-aligned anchor set; any change to the judge is versioned as \textsc{\metric}$_{\textit{model-time}}$ and applied uniformly (Appendix~\ref{sec:version-protocol}).

% \textbf{Cost/latency and accuracy.} Although \metric is reasoning-centric, a Prover-first, early-exit design and concise prompts keep overall cost low (Appendix~\ref{sec:appendix-cost}), and parallel execution renders wall-clock time practical for offline benchmarking (Appendix~\ref{sec:appendix-latency}).

\section{Experiments: Benchmarking}
We conduct a comprehensive re-evaluation on a representative set of NL2SQL methods on two metrics: EX and ROSE$_{\textit{o3-2504}}$.  ROSE in subsequent paragraphs all refers to ROSE$_{\textit{o3-2504}}$.

\subsection{Setup}

\subsubsection{Dataset}
BIRD Mini-Dev is used for evaluation. The split contains 500 NL-SQL pairs from 11 databases. Data remain unchanged compared with BIRD Dev.

\subsubsection{NL2SQL Methods}
Our evaluation considers both prompting methods and fine-tuned methods in NL2SQL. We prioritize systems with publicly accessible prediction outputs on the BIRD leaderboard\footnote{https://bird-bench.github.io/}.

\vspace{2mm}

\noindent\textbf{Prompting Methods}. This category utilizes LLMs via prompt engineering, forgoing task-specific fine-tuning. It can be divided into two sub-categories:

\vspace{-2mm}

\begin{itemize}[leftmargin=4pt, itemsep=0pt]
    \item \textit{Base Models}. We use several LLMs in a zero-shot setting directly. These include OpenAI GPT series (GPT-5, GPT-4o, GPT-4, and GPT-3.5) and DeepSeek-Chat.
    
    \item \textit{Engineered Systems}. This consists of systems that employ multi-step pipelines to decompose the task into sub-problems. The evaluated systems include Alpha-SQL-32B, OpenSearch-SQL, RSL-SQL, SuperSQL, TA-SQL, DAIL-SQL, C3-SQL~\cite{Dong2023C3}, and CoT~\cite{li2023bird}.
\end{itemize}

% \vspace{-2mm}

\noindent\textbf{Fine-tuned Methods}. This category comprises models that are specifically fine-tuned on NL2SQL corpora to specialize their capabilities. The included methods are CSC-SQL, OmniSQL-32B~\cite{Li2025OmniSQL}, CodeS-15B~\cite{Li2024CodeS}, CHESS\textsubscript{(IR,SS,CG)}~\cite{Talaei2024CHESS}, and RESDSQL-3B~\cite{Li2023RESDSQL}.

Details of the engineered systems and fine-tuned methods are in Appendix~\ref{sec:appendix_nl2sql_details}.

\subsection{Insights}

\newcommand{\aclcallout}[1]{%
  \begin{tcolorbox}[
    enhanced,
    breakable,
    colback=blue!3,
    colframe=blue!60!black,
    boxrule=0.6pt, arc=2pt,
    left=3pt,right=3pt,top=3pt,bottom=3pt
  ]
    \textbf{\textcolor{SteelBlue}{#1}}
  \end{tcolorbox}%
}

\subsubsection{Base Model Dominance }
We reveal a foundational principle:

\aclcallout{NL2SQL performance advancement relies largely on the capability of base models rather than engineering designs.}

\begin{figure}[t!]
\centering
\includegraphics[width=1.05\linewidth]{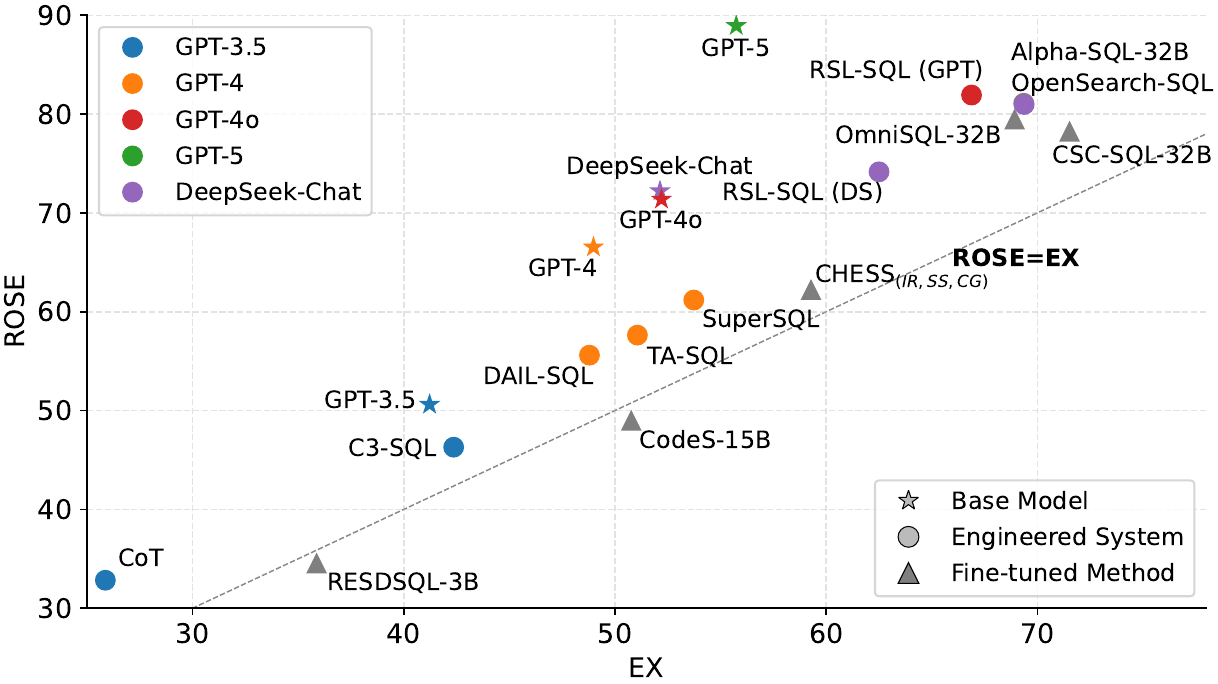}
\caption{\metric vs. EX scores for NL2SQL methods grouped by base model.}
\label{fig:scatter}
\vspace{-5mm}
\end{figure}

This hierarchy is visually evident in Figure~\ref{fig:scatter}, where systems cluster into distinct performance tiers defined by their base model. Performance scales monotonically with model generation, from GPT-3.5, through GPT-4 and GPT-4o, to GPT-5, and engineered systems inherit this ordering. For example, RSL-SQL based on GPT-4o outperforms C3-SQL based on GPT-3.5 in both metrics. This pattern also holds for fine-tuned models: at a similar 30B scale, OmniSQL built on Qwen2.5 surpasses CHESS\textsubscript{(IR,SS,CG)} on DeepSeek-Coder.

We also note that several systems trail their base models (e.g., GPT-4 surpasses DAIL-SQL). This discrepancy arises because our baselines use the latest model, while the systems' public results were generated with earlier versions. This reinforces our finding that recent performance gains are overwhelmingly attributable to stronger base models. It is therefore imperative that future evaluations decouple system design from model updates to accurately attribute true algorithmic improvements beyond model scaling.

\subsubsection{Widening Gap}

Building on the principle, our analysis uncovers a critical paradox: 

\aclcallout{As models become more powerful, the standard metric (EX) serves as an increasingly unreliable indicator of semantic correctness.}

\begin{figure}[t]
\centering
\includegraphics[width=1\linewidth]{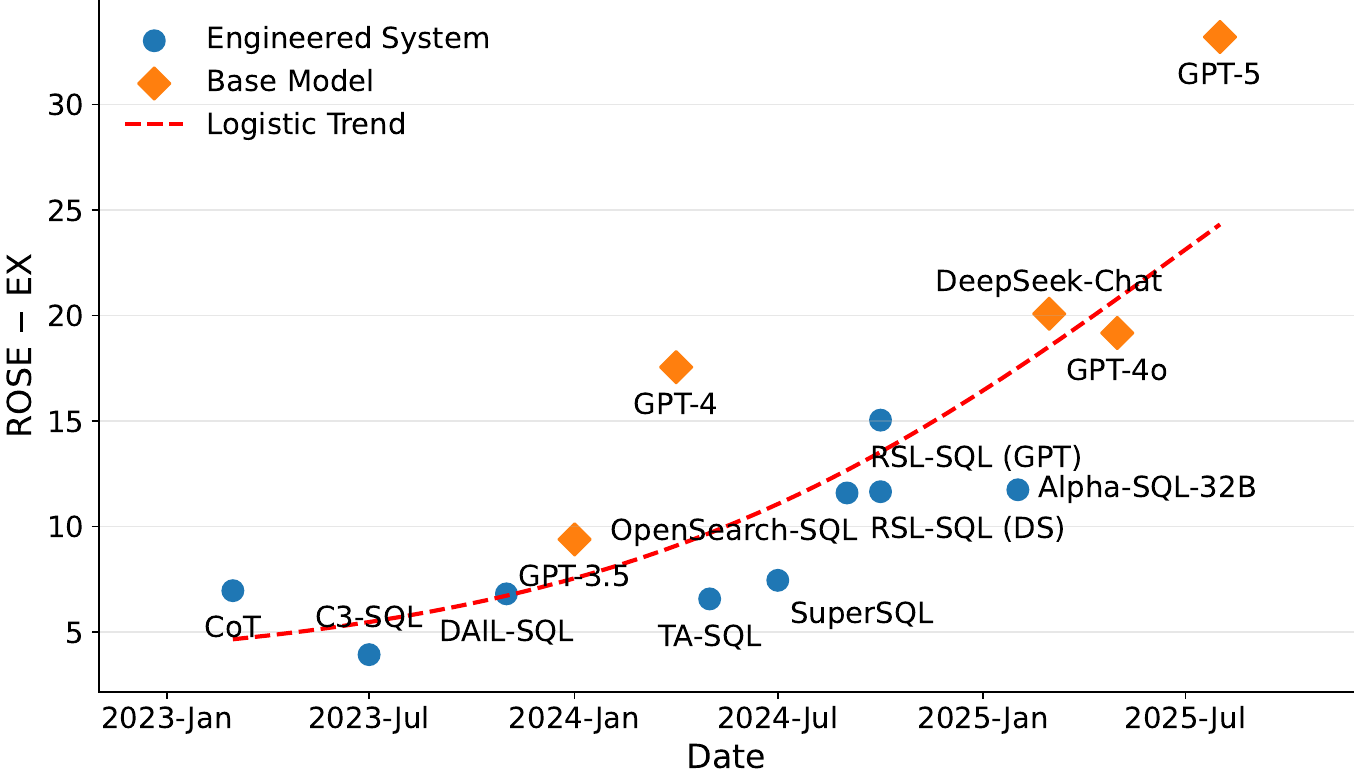}
\caption{The growing divergence between \metric and EX over time for prompting-based systems.}
\label{fig:time}
\vspace{-3mm}
\end{figure}

As illustrated in Figure~\ref{fig:time}, the gap between \metric (intent-centered) and EX (reference-dependent) for prompting methods widens dramatically over time. The trend is stark: the gap was less than 5\% with systems from mid-2023, but balloons to more than 20\% with models projected for mid-2025.

This growing divergence is driven by two compounding factors. First, the nature of errors changes. Early systems often produced semantically incorrect SQL, where both metrics would agree on the failure, keeping the gap small. As methods advance, it is increasingly common to generate SQL that is semantically correct yet flagged wrong by EX due to its shortcomings or erroneous ground-truth SQL.
\metric gives correct judgment for valid solutions in such cases, thus widening the gap. Second, stronger models exhibit greater expressive freedom, generating a wider variety of semantically correct but stylistically diverse SQL. \metric rewards this semantic competence, while EX punishes it, inflating the difference.

This trend delivers a verdict: the community is facing a metric crisis. The widening gap is not merely a statistical artifact, but a clear signal that rigid reference-matching evaluations are becoming obsolete in the era of powerful generative models. This points to the urgent need for an intent-centered metric that can measure true progress in the field.

\subsubsection{Benchmark Flaws}

\vspace{2mm}

\aclcallout{The metric divergence largely stems from benchmark flaws, namely incorrect ground-truth SQL and ambiguous questions.}

To pinpoint these factors, we use \metric's diagnostic labels to isolate the discordance rate on these problematic subsets. The evidence presented in Table~\ref{tab:discordance} is conclusive. For questions with erroneous ground-truth SQL (GoldX), the disagreement rate between the two metrics skyrockets to over 80\% across all tested systems. For ambiguous questions (AmbQ), it remains exceptionally high at around 60\%. Both rates far eclipse the average disagreement of less than 20\% in the overall dataset, confirming that these dataset issues are a significant catalyst for metric divergence.

\begin{table}[hbtp]
\vspace{-1mm}
\centering
\caption{Discordance rate (\%) between EX and \metric on GoldX, AmbQ.}
\label{tab:discordance}
\resizebox{\columnwidth}{!}{%
\begin{tabular}{lcccc}
\toprule
Method & GoldX & AmbQ & GoldX $\cup$ AmbQ & Avg. \\
\midrule
OpenSearch-SQL     & 86.00 & 58.33 & 77.14 & 20.35 \\
Alpha-SQL-32B  & 90.00 & 47.62 & 76.12 & 19.13 \\
RSL-SQL        & 84.62 & 68.18 & 77.94 & 18.08 \\
OmniSQL-32B    & 82.61 & 60.00 & 74.60 & 18.18 \\
\bottomrule
\end{tabular}}
\vspace{-3mm}
\end{table}

\begin{figure}[hbtp]
\centering
\includegraphics[width=\linewidth]{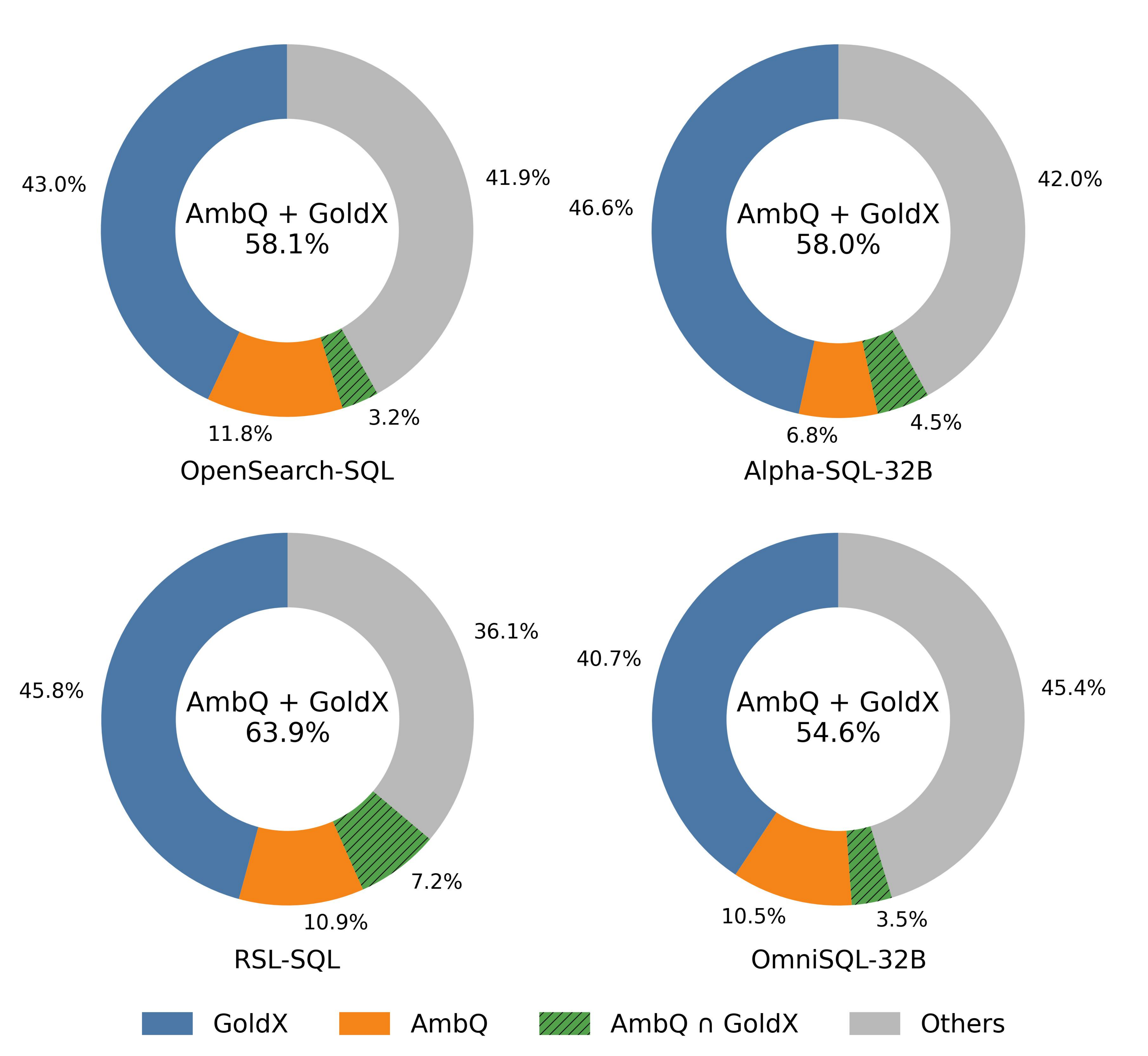}
\caption{Share of discordant cases by question type.}
\label{fig:discordance}
\vspace{-3mm}
\end{figure}

Figure~\ref{fig:discordance} shows that these two culprits are the dominant source of disagreements. GoldX and AmbQ consistently account for roughly 45\% and 10\% of discordant cases, collectively explaining more than half of instances where EX and ROSE diverge. Thus, improving the clarity and correctness of future datasets is essential for making EX a faithful proxy for user intent. This also underscores the critical need to identify benchmark flaws.

\subsubsection{Fine-Tuning Effect}

In contrast, fine-tuning appears to alleviate the widening gap. Our analysis reveals that this is an alignment effect:

\aclcallout{Fine-tuning narrows the metric gap by compelling models to mimic stylistic conventions of datasets.}

\begin{figure}[htbp]
\centering
\includegraphics[width=\linewidth]{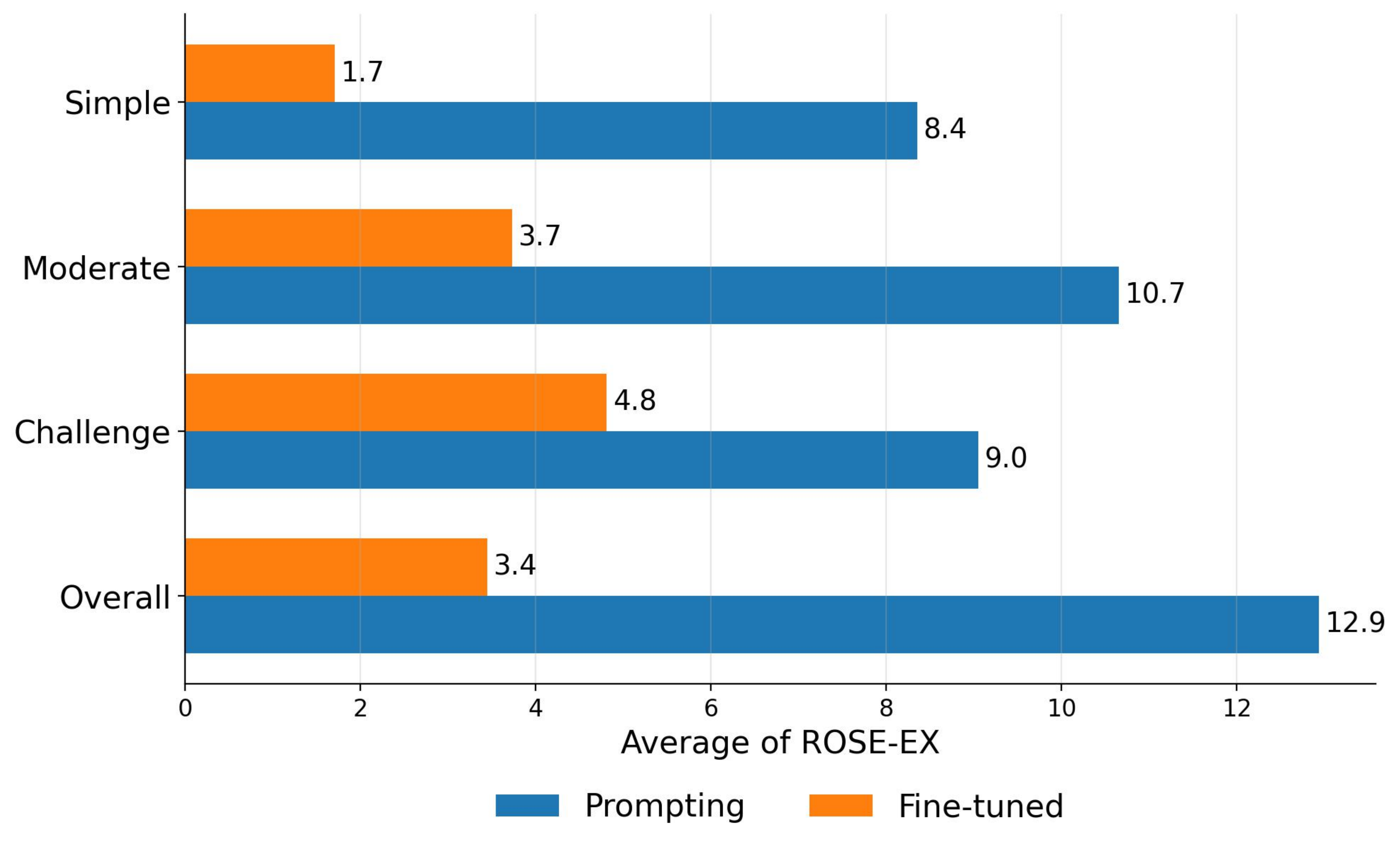}
\caption{Average gap between \metric and EX across difficulty levels for prompting and fine-tuned methods.}
\label{fig:gap}
\vspace{-3mm}
\end{figure}

This phenomenon is quantified in Figure~\ref{fig:gap}, which shows that prompting methods consistently exhibit a larger gap across all difficulty levels. The difference is particularly obvious on Simple and Moderate questions, where the gap for prompting is roughly four times that of fine-tuned methods.

The underlying cause is that most fine-tuned methods are trained on BIRD Train, effectively learning to mimic the SQL style. This is reinforced by the outlier: OmniSQL, primarily trained on a huge corpus (SynSQL), shows a larger metric gap. Conversely, methods such as CodeS and CHESS\textsubscript{(IR,SS,CG)}, which can potentially overfit noisy or limited training data, can even achieve higher EX than \metric.

In contrast, prompting methods lack this dataset-specific conditioning. They generate a wider diversity of semantically correct, yet stylistically varied, SQL. While \metric correctly credits this diversity, EX's rigid reference matching penalizes these variations, which explains the consistently larger gap observed for this class of methods.

\section{Conclusion}
In this work, we addressed the growing crisis in NL2SQL evaluation by introducing \metric, an intent-centered metric that leverages an adversarial Prover-Refuter cascade to achieve superior alignment with expert judgment. We demonstrated its effectiveness on \dataset, our publicly released dataset of 585 expert-annotated samples, where it substantially outperforms existing metrics.

Our re-evaluation yielded four insights for the community. First, the dominance of base models over system engineering calls for a re-evaluation of what constitutes a novel algorithmic contribution. Second, the widening gap between \metric and EX serves as an urgent call to move beyond reference-dependent evaluation. Third, we identified that this gap is primarily caused by addressable benchmark flaws, namely erroneous ground-truth SQL and question ambiguity. Finally, our analysis suggests that fine-tuning on the current training set may be teaching stylistic conformity rather than improving semantic reasoning.

These insights highlight the urgent need for the community to prioritize the development of more robust and intent-centered evaluation metrics and to focus on improving the quality and clarity of datasets. The paradigm pioneered by \metric offers a promising path forward, not only for NL2SQL, but potentially for other complex tasks where success should be defined by fulfilling user intent rather than matching a single, potentially flawed ground-truth solution. To truly measure and guide progress in the era of powerful generative models, we must evolve our methods of evaluation.

\newpage
\section*{Limitations}
While ROSE introduces a more robust, intent-centered evaluation paradigm for NL2SQL, we acknowledge several limitations that warrant consideration for future work.

\vspace{2mm}
\noindent \textbf{Dependency on Foundational LLMs.}
The efficacy of ROSE is intrinsically linked to the reasoning capabilities of its underlying LLM. As demonstrated in Table \ref{tab:verification-metrics}, performance varies when different foundational models are used as the backbone for SQL Prover and Adversarial Refuter. Consequently, the reliability of ROSE as a metric may fluctuate as new models are released. Although such updates create an opportunity for stronger judges that align more closely with expert judgments, they also introduce a risk of drift across versions. Under our version management strategy in Appendix~\ref{sec:version-protocol}, new models are treated as candidates that must pass re-validation before adoption.

\vspace{2mm}
\noindent \textbf{Selection Bias from Consensus Filtering.}
ROSE-VEC is constructed by retaining only cases on which two annotators reach exact agreement. This reduces label noise and yields a high-confidence validation set, but it may also introduce a selection bias. In particular, retained instances may over-represent queries with clearer interpretation or easier semantic judgments, while under-representing borderline, disagreement-prone, or genuinely ambiguous cases. As a result, the reported agreement between automatic metrics and expert labels may not fully reflect performance on the broader distribution of NL2SQL outputs.

\vspace{2mm}
\noindent \textbf{Computational Cost and Latency.}
Compared to deterministic metrics such as Execution Accuracy (EX), LLM-based metrics like ROSE are substantially more resource-intensive. The computational overhead and increased latency associated with model inference may limit the practicality of using ROSE in scenarios requiring rapid, iterative method development and testing, where immediate feedback is crucial. ROSE mitigates this overhead through call routing, concise prompts, and multi-thread parallelism, but it remains more expensive than deterministic alternatives.

\section*{Acknowledgements}
This paper was supported by the NSF of China (62402409); Youth S\&T Talent Support Programme of Guangdong Provincial Association for Science and Technology (SKXRC2025461); the Young Talent Support Project of Guangzhou Association for Science and Technology (QT-2025-001); Guangzhou Basic and Applied Basic Research Foundation (2026A1515010269, 2025A04J3935, 2023A1515110545); and Guangzhou-HKUST(GZ) Joint Funding Program (2025A03J3714).

\bibliography{custom}

\clearpage
\appendix

\section{Mathematical Definitions of NL2SQL Metrics}
\label{sec:math-nl2sql-metrics}
\begin{itemize}
    \item \textbf{Exact Match (EM).} Given a predicted SQL $S_p$ and a ground-truth SQL $S_g$, with their normalized forms denoted as $\hat{S}_p$ and $\hat{S}_g$ respectively:
    \[
    \text{EM}(S_p, S_g) = (\hat{S}_p \equiv \hat{S}_g)
    \]

    \item \textbf{Component Match (CM).} The score is the proportion of the components from the ground-truth SQL $S_g$ that match the corresponding components in the predicted query $S_p$:
    \[
    \text{CM}(S_p, S_g) = \frac{\sum_{c \in C(S_g)} (c_p \cong c)}{|C(S_g)|}
    \]
    where $C(S_g)$ is the set of components in $S_g$, $c_p$ is the corresponding component in $S_p$, and $\cong$ denotes a successful match.

    \item \textbf{Execution Accuracy (EX).} A prediction is correct if its execution result set $E_p$ is a multiset equivalent of the ground-truth result set $E_g$ when run on a database $D$:
    \[
    \text{EX}(S_p, S_g, D) = (E_p \equiv E_g)
    \]

    \item \textbf{Enhanced Tree Match (ETM).} A match is declared if the canonical forms of the queries' Abstract Syntax Trees (ASTs), denoted as $\widehat{\mathrm{AST}}$, are structurally equivalent:
    \[
    \text{ETM}(S_p, S_g, D) = (\widehat{\mathrm{AST}}(S_p) \equiv \widehat{\mathrm{AST}}(S_g))
    \]

    \item \textbf{LLM-based Metric.} The score is the output of an LLM prompted with the relevant context, where $Q$ is the natural language question, $D$ is the database schema, and $C$ represents user- or task-specific acceptance criteria:
    \[
    \text{Score} = \text{LLM}(\text{prompt}(Q, S_p, S_g, E_p, E_g \ |D, C))
    \]
\end{itemize}

\section{Mathematical Definition of ROSE}
\label{sec:math-rose}

$$
\text{\metric} =
\begin{cases}
0 & \text{if not } \sigma_{\text{syn}}(S_p\ |D) \\
0 & \text{if not } j_{p} \\
0 & \text{if } j_{r} \\
1 & \text{otherwise}
\end{cases}
$$

\section{Comparison of Evaluation Metrics}
\label{sec:appendix_comparison}
To summarize our contributions, Table \ref{tab:metric_comparison} outlines the key differences between ROSE and prior evaluation metrics. The comparison highlights how existing methods are fundamentally limited by their reliance on a single ground-truth SQL, whereas ROSE is designed for greater robustness against ambiguity and ground-truth errors, offering diagnostic capabilities beyond a simple binary score.

\begin{table*}[!ht]
\centering
\caption{Key differences between our proposed metric, ROSE, and prior evaluation metrics.}
\label{tab:metric_comparison}
{
\setlength{\tabcolsep}{2pt}
\resizebox{\linewidth}{!}{%
\begin{tabular}{lcccc}
\toprule
\textbf{Aspect} & \textbf{EX} & \textbf{LLM-SQL-Solver} & \textbf{FLEX} & \textbf{ROSE} \\
\midrule
Basis for Judgment & Match & Match & Comparison & Independent and Comparison \\
\addlinespace
Role of Ground-truth SQL & Reference & Reference & Reference & Challenge Evidence \\
\addlinespace
Robust to Ambiguity & No & No & Partially & Yes \\
\addlinespace
Robust to Ground-truth Error & No & No & Partially & Yes \\
\addlinespace
Diagnostic Output & No & No & No & Yes (\texttt{AmbQ}, \texttt{GoldX}) \\
\bottomrule
\end{tabular}%
}
}
\end{table*}

\section{Formulations of Validation Metrics}
\label{sec:appendix_metrics}

This section provides the mathematical definitions for the validation metrics used in Section 5.1.2. Let TP, FP, TN, and FN denote the counts of true positives, false positives, true negatives, and false negatives from the confusion matrix, respectively.

\begin{itemize}[leftmargin=*]
    \item \textbf{Accuracy} is the ratio of correct predictions to the total number of predictions.
    \[
    \mathrm{Acc}=\frac{TP+TN}{TP+TN+FP+FN}
    \]

    \item \textbf{Cohen's Kappa} measures inter-rater agreement for categorical items.
    \[
    \kappa=\frac{p_o-p_e}{1-p_e}
    \]
    Here, $p_o=(TP+TN)/N$ is the observed agreement, and $p_e=\pi_1\rho_1+\pi_0\rho_0$ is the chance agreement from the marginals, with $\pi_1=(TP+FP)/N$, $\rho_1=(TP+FN)/N$, $\pi_0=1-\pi_1$, and $\rho_0=1-\rho_1$, where $N$ is the total number of samples.

    \item \textbf{MCC (Matthews Correlation Coefficient)} is a correlation coefficient between the observed and predicted binary classifications.
    \[
    \resizebox{0.9\linewidth}{!}{$
    \mathrm{MCC}=\frac{TP\cdot TN-FP\cdot FN}{\sqrt{(TP+FP)(TP+FN)(TN+FP)(TN+FN)}}
    $}
    \]

    \item \textbf{F1-score} is the harmonic mean of precision and recall.
    \[
    \mathrm{F1}=\frac{2 \cdot P \cdot R}{P+R}=\frac{2TP}{2TP+FP+FN}
    \]
    Here, $P=TP/(TP+FP)$ is precision and $R=TP/(TP+FN)$ is recall.
\end{itemize}

\section{Version Management}
\label{sec:version-protocol}

\subsection{Naming Convention}
\label{sec:naming-convention}
We denote each instantiated \metric judge as \textsc{\metric}$_{\textit{model-time}}$, where \textit{model} is the backbone identifier and \textit{time} is the backbone release month in \textit{yymm} format. For example, \textsc{\metric}$_{\textit{o3-2504}}$ refers to the o3 backbone released in April 2025. All reported scores should include this version tag to make the judge configuration explicit.

\subsection{Backbone Versioning}
\label{sec:backbone-versioning}
LLM-based metrics are sensitive to the choice of backbone models. Because foundation models evolve rapidly, frequent model updates can introduce drift in reported scores. However, they also create an opportunity for improved evaluation when stronger reasoning capabilities yield better alignment with expert judgments. To ensure fair comparison for public leaderboards, we commit to a unified backbone within a given period.

The unified backbone is selected among candidate backbones by validating agreement against ROSE-VEC. We replace the unified backbone if and only if a new candidate achieves better validation performance on ROSE-VEC across all four metrics: Accuracy, Cohen's $\kappa$, MCC, and F1. This requirement promotes principled upgrades while preventing regressions in any individual aspect.

When the default judge is updated, earlier scores remain valid and are retained, but are tagged with their original \textsc{ROSE}$_{\textit{model-time}}$ version. New evaluations are published under the new version tag. In this way, judge changes are controlled, documented, and versioned.

\section{Additional Ablation Studies}
\label{sec:appendix-ablations}

We perform additional ablations on ROSE-VEC-BIRD to study two design choices of \metric: (i) whether the judge has access to ground-truth SQL, and (ii) whether verification and critique are carried out in a single unified prompt or decomposed. We run ablations on two reasoning backbones, OpenAI o3-2504 and Gemini-2.5 Pro-2506.

Table~\ref{tab:rose-ablation} reports the results. Across both backbones, all ablated variants fall behind the full ROSE cascade, showing that removing ground-truth supervision or collapsing two stages into a single prompt both degrade agreement with expert judges. This confirms that access to reference SQL and multi-stage Prover-Refuter reasoning are both important components of \metric.

\section{Other Open-Source Backbones}
\label{sec:appendix-opensource}

We further evaluate three open-source reasoning backbones on the BIRD split of ROSE-VEC (ROSE-VEC-BIRD): Qwen3-235B-A22B-Thinking-2507, Qwen3-30B-A3B-Thinking-2507, and DeepSeek-R1-Distill-Qwen-32B-2501. Table~\ref{tab:rose-vec-bird-opensource} reports EX as a deterministic baseline and compares ROSE with FLEX under each backbone. ROSE consistently improves over FLEX across all four validation metrics. These results show that ROSE transfers well to smaller and open-source models.

\begin{table*}[hbt]
\centering
\begingroup
\fontsize{10pt}{12pt}\selectfont
\setlength{\tabcolsep}{14pt}
\caption{Performance on ROSE-VEC-BIRD with other open-source backbones.}
\label{tab:rose-vec-bird-opensource}
\begin{tabular}{l l r r r r}
\toprule
\textbf{Backbone} & \textbf{Metric} & \textbf{\(\boldsymbol{\kappa}\) (\%)} & \textbf{Acc (\%)} & \textbf{MCC (\%)} & \textbf{F1 (\%)} \\
\midrule
\multirow{1}{*}{\textbf{Deterministic}}
& EX & 43.56 & 69.57 & 51.36 & 71.18 \\
\midrule
\multirow{2}{*}{\textbf{\begin{tabular}[c]{@{}l@{}}Qwen3-235B\\-A22B-Thinking\end{tabular}}}
& FLEX        & 53.59 & 76.71 & 54.89 & 74.40 \\
& \textbf{ROSE} & \textbf{65.11} & \textbf{83.85} & \textbf{65.71} & \textbf{87.38} \\
\midrule
\multirow{2}{*}{\textbf{\begin{tabular}[c]{@{}l@{}}Qwen3-30B\\-A3B-Thinking\end{tabular}}}
& FLEX        & 47.74 & 72.98 & 52.78 & 75.90 \\
& \textbf{ROSE} & \textbf{62.63} & \textbf{81.99} & \textbf{64.59} & \textbf{85.28} \\
\midrule
\multirow{2}{*}{\textbf{\begin{tabular}[c]{@{}l@{}}DeepSeek-R1-Distill\\-Qwen-32B\end{tabular}}}
& FLEX        & 43.01 & 72.36 & 43.33 & 66.16 \\
& \textbf{ROSE} & \textbf{56.86} & \textbf{80.12} & \textbf{57.28} & \textbf{84.54} \\
\bottomrule
\end{tabular}
\endgroup
\end{table*}

\begin{table*}[hbt]
\centering
\begingroup
\fontsize{10pt}{12pt}\selectfont
\setlength{\tabcolsep}{14pt}
\caption{Ablation studies of \metric on ROSE-VEC-BIRD. “w/o GT” indicates  removing ground-truth supervision.}
\label{tab:rose-ablation}
\begin{tabular}{l l r r r r}
\toprule
\textbf{Backbone} & \textbf{Metric} & \textbf{\(\boldsymbol{\kappa}\) (\%)} & \textbf{Acc (\%)} & \textbf{MCC (\%)} & \textbf{F1 (\%)} \\
\midrule
\multirow{4}{*}{\textbf{OpenAI o3}}
& Unified w/o GT   & 53.35 & 80.43 & 54.00 & 86.09 \\
& Unified          & 66.35 & 83.85 & 68.22 & 86.87 \\
& ROSE w/o GT      & 71.01 & 86.34 & 72.25 & 89.11 \\
& \textbf{ROSE}    & \textbf{80.68} & \textbf{90.99} & \textbf{81.64} & \textbf{92.91} \\
\midrule
\multirow{4}{*}{\textbf{Gemini-2.5 Pro}}
& Unified w/o GT   & 45.06 & 78.88 & 50.07 & 86.01 \\
& Unified          & 59.90 & 81.06 & 61.02 & 84.86 \\
& ROSE w/o GT      & 54.94 & 80.43 & 55.01 & 85.65 \\
& \textbf{ROSE}    & \textbf{64.79} & \textbf{82.92} & \textbf{67.15} & \textbf{85.93} \\
\bottomrule
\end{tabular}
\endgroup
\end{table*}

\section{Efficiency}
\subsection{Time Efficiency}
\label{sec:appendix-latency}
\begin{table}[hbt]
\centering
\begingroup
\fontsize{10pt}{12pt}\selectfont
\setlength{\tabcolsep}{14pt}
\caption{Wall-clock time under different threads.}
\label{tab:rose-latency-wallclock}
\begin{tabular}{r r}
\toprule
\textbf{Threads} & \textbf{Wall-clock Time} \\
\midrule
1 & 120min37s \\
2 & 64min59s \\
4 & 42min41s \\
8 & 18min \\
\bottomrule
\end{tabular}
\endgroup
\end{table}

We measure the wall-clock time of ROSE$_{\textit{o3-2504}}$ on ROSE-VEC-BIRD under different numbers of threads. As shown in Table~\ref{tab:rose-latency-wallclock}, parallel execution substantially reduces end-to-end evaluation time.

\begin{table}[hbt]
\centering
\begingroup
\fontsize{10pt}{12pt}\selectfont
\setlength{\tabcolsep}{14pt}
\caption{Average time on ROSE-VEC-BIRD.}
\label{tab:rose-latency-avg}
\begin{tabular}{l r}
\toprule
\textbf{Metric} & \textbf{Avg.\ Time} \\
\midrule
EX & 1.22s \\
FLEX$_{\textit{o3-2504}}$ (thread-1) & 18.19s \\
ROSE$_{\textit{o3-2504}}$ (thread-1) & 22.48s \\
ROSE$_{\textit{o3-2504}}$ (thread-8) & 3.35s \\
\bottomrule
\end{tabular}
\endgroup
\end{table}

Table~\ref{tab:rose-latency-avg} further reports the average time per question. On a single thread, ROSE incurs higher per-question latency than FLEX, but multi-threading significantly lowers the effective per-question time, making ROSE practical for benchmarking settings where parallel execution is available.

\subsection{Cost Efficiency}
\label{sec:appendix-cost}

We compare the monetary cost of ROSE$_{\textit{o3-2504}}$ and FLEX$_{\textit{o3-2504}}$ on ROSE-VEC-BIRD. Although ROSE may invoke a second-stage Refuter, it remains cost-efficient in practice due to concise prompts and conditional second-stage calls. Table~\ref{tab:rose-cost} reports both total cost and average cost per query.

\begin{table}[ht]
\centering
\begingroup
\fontsize{10pt}{12pt}\selectfont
\setlength{\tabcolsep}{4pt}
\caption{Monetary cost on ROSE-VEC-BIRD.}
\label{tab:rose-cost}
\begin{tabular}{l r r}
\toprule
\textbf{Metric} & \textbf{Total Cost (USD)} & \textbf{Avg.\ Cost (USD)} \\
\midrule
FLEX$_{\textit{o3-2504}}$ & 3.819 & 0.0118 \\
ROSE$_{\textit{o3-2504}}$ & 2.249 & 0.0070 \\
\bottomrule
\end{tabular}
\endgroup
\end{table}

To explain the cost behavior, we report the number of LLM calls used by ROSE$_{\textit{o3-2504}}$ on ROSE-VEC-BIRD. As shown in Table~\ref{tab:rose-calls}, more than half of the questions are completed with a single call, and the average number of calls per question is 1.45.

\begin{table}[ht]
\centering
\begingroup
\fontsize{10pt}{12pt}\selectfont
\setlength{\tabcolsep}{10pt}
\caption{Number of LLM calls by ROSE$_{\textit{o3-2504}}$ on ROSE-VEC-BIRD.}
\label{tab:rose-calls}
\begin{tabular}{r r r}
\toprule
\textbf{\#Calls} & \textbf{\#Questions} & \textbf{Percentage (\%)} \\
\midrule
1 & 176 & 54.66 \\
2 & 146 & 45.34 \\
\midrule
\textbf{Avg.} & \textbf{1.45} & -- \\
\bottomrule
\end{tabular}
\endgroup
\end{table}

\section{Error Analysis}
\label{sec:error-analysis}

% Optional: slightly increase spacing for the example blocks only.
\newcommand{\exsep}{\vspace{0.35em}}
\newcommand{\sqlstyle}[1]{{\small\texttt{#1}}}

To better understand when and why ROSE fails, we audit all disagreement instances between ROSE and expert labels on ROSE-VEC-BIRD using \textsc{ROSE}$_{\textit{o3-2504}}$. Among 29 failure cases, 26 are false negatives and 3 are false positives (Q252, Q861, Q1487), suggesting that ROSE is generally conservative and more likely to under-credit than to over-credit. We summarize the observed failure modes below.

\subsection{Logical Robustness vs.\ Coincidental Correctness}
This is the dominant pattern, accounting for roughly half of the disagreements (15/29). ROSE penalizes SQL that happens to return the correct answer on the current database state but is logically fragile under plausible data variations.

\exsep
\noindent\textbf{Example (Q137).}\par
\exsep
\noindent\textbf{Question.} ``How many accounts have running contracts in Branch location 1?''\par
\exsep
\noindent\textbf{Predicted SQL.} \sqlstyle{SELECT COUNT(account.account\_id) FROM account INNER JOIN loan ON account.account\_id = loan.account\_id WHERE account.district\_id = 1 AND (loan.status = 'C' OR loan.status = 'D');}\par
\exsep
\noindent\textbf{Analysis.} The query counts loan rows rather than distinct accounts. It matches the gold answer on the current instance, but it would over-count once an account can hold multiple contracts. ROSE flags this as logically unsafe even though execution coincides with the reference.\par
\exsep

\subsection{Pragmatic Rigidity in Units and Format}
Around one third of the disagreements (11/29) arise from strictness about representation, including units, scaling, and output conventions. Annotators may accept a representation as equivalent, while ROSE treats it as semantically incomplete.

\exsep
\noindent\textbf{Example (Q77).}\par
\exsep
\noindent\textbf{Question.} ``Which schools served a grade span of Kindergarten to 9th grade in the county of Los Angeles and what is its Percent (\%) Eligible FRPM (Ages 5--17)?''\par
\exsep
\noindent\textbf{Predicted SQL.} \sqlstyle{SELECT schools.School, frpm."Percent (\%) Eligible FRPM (Ages 5-17)" FROM schools INNER JOIN frpm ON schools.CDSCode = frpm.CDSCode WHERE schools.GSserved = 'K-9' AND schools.County = 'Los Angeles';}\par
\exsep
\noindent\textbf{Analysis.} The query returns a stored FRPM value represented as a proportion in $[0,1]$, while the question requests a percentage. Annotators accept the proportional form up to a deterministic scaling factor. ROSE treats the unit mismatch as a semantic gap and marks it incorrect.\par
\exsep

\subsection{World Knowledge, Schema Heuristics, and Ambiguity}
The remaining cases (3/29) are driven by domain interpretation, schema heuristics, or genuine ambiguity in the natural-language query, where multiple readings can be defensible and ROSE may enforce a more literal interpretation than annotators.

\exsep
\noindent\textbf{Example (Q82).}\par
\exsep
\noindent\textbf{Question.} ``What is the grade span offered in the school with the highest longitude?''\par
\exsep
\noindent\textbf{Predicted SQL.} \sqlstyle{SELECT GSoffered FROM schools ORDER BY ABS(Longitude) DESC LIMIT 1;}\par
\exsep
\noindent\textbf{Analysis.} The query orders by absolute longitude, effectively selecting the school farthest from the prime meridian. Annotators consider this a plausible reading of ``highest longitude''. ROSE enforces the numerically largest longitude and marks the prediction as incorrect.\par
\exsep

\begin{table}[hbtp]
\begingroup
\fontsize{10pt}{11pt}\selectfont
\setlength{\tabcolsep}{6pt}
\renewcommand{\arraystretch}{1.2}
\caption{Precision of \metric diagnostic labelling for GoldX and AmbQ on ROSE-VEC-BIRD.}
\label{tab:verification-tags-bird}
\begin{tabular}{l l r r}
\toprule
\textbf{Label} & \textbf{Model} & \textbf{Count} & \textbf{Precision (\%)} \\
\midrule
\multirow{3}{*}{GoldX}
& OpenAI o3        & 35 & \textbf{82.86} \\
& Gemini-2.5 Pro   & 58 & 63.79 \\
& DeepSeek-R1      & 59 & 57.63 \\
\midrule
\multirow{3}{*}{AmbQ}
& OpenAI o3        & 16 & \textbf{75.00} \\
& Gemini-2.5 Pro   & 17 & 35.29 \\
& DeepSeek-R1      & 61 & 27.87 \\
\bottomrule
\end{tabular}
\endgroup
\vspace{-5mm}
\end{table}

\begin{table*}[hbt]
\centering
\begingroup
\fontsize{10pt}{12pt}\selectfont
\setlength{\tabcolsep}{14pt}
\renewcommand{\arraystretch}{1.15}
\caption{Performance of NL2SQL metrics on ROSE-VEC-BIRD. \textbf{Bold} indicates the best metric within each LLM block. \underline{Underline} indicates the second best within each block. Open-source models are represented with {\small\unlockicon{}}, while closed-source models are represented with {\small\lockicon{}}.}
\label{tab:verification-metrics-bird}
\begin{tabular}{l l r r r r}
\toprule
\textbf{Backbone} & \textbf{Metric} & \textbf{\(\boldsymbol{\kappa}\) (\%)} & \textbf{Acc (\%)} & \textbf{MCC (\%)} & \textbf{F1 (\%)} \\
\midrule
\multirow{3}{*}{\textbf{Deterministic}}
& EM   & 0.00  & 32.73 & 0.00  & 0.00  \\
& ETM  & \underline{8.60}  & \underline{41.30} & \underline{21.19} & \underline{22.22} \\
& EX   & \textbf{43.56} & \textbf{69.57} & \textbf{51.36} & \textbf{71.18} \\
\midrule
\multirow{4}{*}{\textbf{OpenAI o3}\textsuperscript{\lockicon}}
& LLM-SQL-Solver & 10.51 & 43.65 & 23.56 & 25.75 \\
& FLEX           & 55.20 & 76.87 & 60.53 & 79.18 \\
& \ablation & \underline{66.76} & \underline{85.67} & \underline{67.25} & \underline{89.57} \\
& \textbf{ROSE}  & \textbf{80.68} & \textbf{90.99} & \textbf{81.64} & \textbf{92.91} \\
\midrule
\multirow{4}{*}{\textbf{Gemini-2.5 Pro}\textsuperscript{\lockicon}}
& LLM-SQL-Solver & 16.55 & 48.45 & 30.03 & 37.59 \\
& FLEX           & \underline{57.68} & 78.98 & \underline{62.20} & 81.40 \\
& \ablation & 51.25 & \underline{80.75} & 54.62 & \textbf{86.97} \\
& \textbf{ROSE}  & \textbf{64.79} & \textbf{82.92} & \textbf{67.15} & \underline{85.93} \\
\midrule
\multirow{4}{*}{\textbf{DeepSeek-R1}\textsuperscript{\unlockicon}}
& LLM-SQL-Solver & 15.02 & 48.14 & 25.56 & 38.83 \\
& FLEX           & \underline{53.87} & 76.40 & \underline{58.80} & 79.23 \\
& \ablation & 52.77 & \underline{79.19} & 52.77 & \underline{84.53} \\
& \textbf{ROSE}  & \textbf{62.13} & \textbf{81.99} & \textbf{63.53} & \textbf{85.50} \\
\bottomrule
\end{tabular}
\endgroup
\end{table*}

\begin{table*}[hbt]
\centering
\begingroup
\fontsize{10pt}{12pt}\selectfont
\setlength{\tabcolsep}{14pt}
\renewcommand{\arraystretch}{1.15}
\caption{Performance of NL2SQL metrics on ROSE-VEC-Spider. \textbf{Bold} indicates the best metric within each LLM block. \underline{Underline} indicates the second best within each block. Open-source models are represented with {\small\unlockicon{}}, while closed-source models are represented with {\small\lockicon{}}.}
\label{tab:verification-metrics-spider}
\begin{tabular}{l l r r r r}
\toprule
\textbf{Backbone} & \textbf{Metric} & \textbf{\(\boldsymbol{\kappa}\) (\%)} & \textbf{Acc (\%)} & \textbf{MCC (\%)} & \textbf{F1 (\%)} \\
\midrule
\multirow{3}{*}{\textbf{Deterministic}}
& EM   & 0.77  & 21.59 & 6.22  & 3.72  \\
& ETM  & \underline{4.47}  & \underline{28.41} & \underline{15.11} & \underline{18.88} \\
& EX   & \textbf{11.09} & \textbf{39.02} & \textbf{24.23} & \textbf{38.31} \\
\midrule
\multirow{4}{*}{\textbf{OpenAI o3}\textsuperscript{\lockicon}}
& LLM-SQL-Solver & 13.06 & 41.83 & 26.43 & 42.70 \\
& FLEX           & \underline{57.85} & 82.13 & \underline{63.23} & 87.47 \\
& \ablation & 45.87 & \underline{84.79} & 47.29 & \underline{90.87} \\
& \textbf{ROSE}  & \textbf{78.61} & \textbf{92.78} & \textbf{78.85} & \textbf{95.40} \\
\midrule
\multirow{4}{*}{\textbf{Gemini-2.5 Pro}\textsuperscript{\lockicon}}
& LLM-SQL-Solver & 21.85 & 52.85 & 35.02 & 58.11 \\
& FLEX           & 28.34 & 60.84 & 39.09 & 67.91 \\
& \ablation & \underline{40.16} & \underline{84.41} & \underline{43.70} & \underline{90.83} \\
& \textbf{ROSE}  & \textbf{75.40} & \textbf{91.63} & \textbf{75.70} & \textbf{94.66} \\
\midrule
\multirow{4}{*}{\textbf{DeepSeek-R1}\textsuperscript{\unlockicon}}
& LLM-SQL-Solver &  13.53 & 44.11 & 25.02 & 46.93 \\
& FLEX           & 10.67 & 38.40 & 23.74 & 37.21 \\
& \ablation & \underline{33.61} & \underline{79.09} & \underline{33.64} & \underline{87.00} \\
& \textbf{ROSE}  & \textbf{66.03} & \textbf{87.63} & \textbf{67.00} & \textbf{92.08} \\
\bottomrule
\end{tabular}
\endgroup
\end{table*}

\section{Results on ROSE-VEC-BIRD and ROSE-VEC-Spider}

\subsection{Metric effectiveness}
\label{sec:appendix_metric_effectiveness}

We report disaggregated results for ROSE-VEC-BIRD in Table~\ref{tab:verification-metrics-bird} and ROSE-VEC-Spider in Table~\ref{tab:verification-metrics-spider}. On both subsets, deterministic metrics lag well behind LLM-based judges, with EX consistently stronger than EM and ETM but still far from expert-level agreement. Within each backbone, ROSE achieves the best agreement and correlation on both splits. The comparison between \ablation and ROSE highlights the contribution of the refuter stage: ROSE improves $\kappa$ and MCC on both subsets, especially on ROSE-VEC-Spider where deterministic signals are weakest. The only exception is ROSE-VEC-BIRD under Gemini-2.5 Pro, where \ablation attains slightly higher F1 while ROSE maintains superior agreement-oriented scores.

\subsection{Diagnostic effectiveness}
\label{sec:appendix_diag_precision}

Tables~\ref{tab:verification-tags-bird} and \ref{tab:verification-tags-spider} report the precision of \metric diagnostic labelling on the BIRD and Spider splits. On both splits, OpenAI o3 attains the highest precision for GoldX and AmbQ, with Gemini-2.5 Pro generally second and DeepSeek-R1 trailing. AmbQ precision is much lower on BIRD for Gemini-2.5 Pro and DeepSeek-R1, indicating that ambiguous-question detection is challenging in this domain. On Spider, all three backbones achieve higher precision and AmbQ precision rises sharply, with o3 reaching 97.56\%. Overall, these results show that \metric yields reliable diagnostic tags when coupled with strong reasoning backbones and that the Spider split provides cleaner conditions for diagnostic labelling than the BIRD split.

\begin{table}[hbtp]
\begingroup
\fontsize{10pt}{11pt}\selectfont
\setlength{\tabcolsep}{6pt}
\renewcommand{\arraystretch}{1.2}
\caption{Precision of \metric diagnostic labelling for GoldX and AmbQ on ROSE-VEC-Spider.}
\label{tab:verification-tags-spider}
\begin{tabular}{l l r r}
\toprule
\textbf{Label} & \textbf{Model} & \textbf{Count} & \textbf{Precision (\%)} \\
\midrule
\multirow{3}{*}{GoldX}
& OpenAI o3        & 16 & \textbf{87.50} \\
& Gemini-2.5 Pro   & 18 & 83.33 \\
& DeepSeek-R1      & 17 & 76.47 \\
\midrule
\multirow{3}{*}{AmbQ}
& OpenAI o3        & 41 & \textbf{97.56} \\
& Gemini-2.5 Pro   & 46 & 86.96 \\
& DeepSeek-R1      & 47 & 82.98 \\
\bottomrule
\end{tabular}
\endgroup
\vspace{-5mm}
\end{table}

\begin{table*}[hbt]
\centering
\caption{Score gap on \textsc{BIRD} Mini-Dev across difficulty levels. \underline{Underlined} methods denote base models.}
\label{tab:score_gap_details}
\resizebox{0.8\linewidth}{!}{%
\begin{tabular}{llcccc}
\toprule
\textbf{Method} & \textbf{Model} & \textbf{Simple} & \textbf{Moderate} & \textbf{Challenge} & \textbf{Overall} \\
\midrule
\multicolumn{6}{c}{\textbf{Prompting Methods}} \\
\midrule
\underline{GPT-5} & -- & 21.24 & 36.21 & 43.44 & 33.19 \\
Alpha-SQL-32B & Qwen2.5-Coder & 10.22 & 10.18 & 17.52 & 11.74 \\
OpenSearch-SQL & DeepSeek-Chat & 13.24 & 10.66 & 11.46 & 11.59 \\
RSL-SQL & DeepSeek-Chat & 12.14 & 11.96 & 10.20 & 11.65 \\
\underline{DeepSeek-Chat} & -- & 11.65 & 23.99 & 22.78 & 20.08 \\
RSL-SQL & GPT-4o & 11.03 & 15.49 & 19.59 & 15.04 \\
\underline{GPT-4o} & -- & 10.53 & 20.61 & 4.95 & 19.17 \\
SuperSQL & GPT-4 & 5.14 & 20.72 & 3.13 & 7.45 \\
TA-SQL & GPT-4 & 5.07 & 5.51 & 11.23 & 6.57 \\
DAIL-SQL & GPT-4 & 9.56 & 8.04 & 0.00 & 6.81 \\
\underline{GPT-4} & -- & 10.28 & 23.87 & 12.87 & 17.55 \\
CoT & GPT-3.5 & 7.35 & 7.02 & 6.25 & 6.96 \\
C3-SQL & GPT-3.5 & 1.47 & 6.22 & 2.06 & 3.93 \\
\underline{GPT-3.5} & -- & 12.33 & 7.82 & 8.91 & 9.39 \\
\midrule
\multicolumn{6}{c}{\textbf{Fine-tuned Methods}} \\
\midrule
CSC-SQL-32B & XiYan-Qwen2.5 & 2.14 & 6.35 & 14.29 & 6.75 \\
OmniSQL-32B & Qwen2.5-Coder & 6.43 & 11.07 & 15.31 & 10.57 \\
CodeS-15B & StarCoder & 2.19 & $-2.21$ & $-6.52$ & $-1.76$ \\
CHESS\textsubscript{(IR,SS,CG)} & DeepSeek-Coder & 0.00 & 4.64 & 3.06 & 2.96 \\
RESDSQL-3B & T5 & $-2.23$ & $-0.44$ & $-2.09$ & $-1.30$ \\
\bottomrule
\end{tabular}%
}
\end{table*}

\section{Details of NL2SQL Methods}
\label{sec:appendix_nl2sql_details}

We provide further details on the advanced engineered systems and fine-tuned methods evaluated in our work.

\subsection{Engineered Systems}
These systems utilize advanced LLMs and apply sophisticated prompt engineering to break down the main task into more manageable sub-problems, such as schema linking, candidate generation, and SQL revision~\cite{hou2026nl2sqlbench}.

\begin{itemize}[leftmargin=*]
    \item \textbf{Alpha-SQL-32B} \cite{li2025alphasql} employs a Monte Carlo Tree Search algorithm to explore the space of partial SQL queries, guide action proposals, and rank final candidates using a search-derived reward signal.
    
    \item \textbf{OpenSearch-SQL} \cite{xie2025opensearchsql} dynamically retrieves few-shot exemplars from a database, generates multiple SQL candidates, and aligns their intermediate representations to enforce cross-candidate consistency.
    
    \item \textbf{RSL-SQL} \cite{cao2024rslsql} enhances the schema linking phase through bidirectional matching and contextual augmentation. It further refines the generated SQL via multi-round self-correction guided by execution feedback.
    
    \item \textbf{SuperSQL} \cite{Li2024SuperSQL} conducts a search over a prompt design space to construct a robust pipeline featuring explicit schema grounding, controlled decoding, and a selective self-consistency voting mechanism.
    
    \item \textbf{TA-SQL} \cite{qu2024tasql} first aligns the user's question to the target database schema to mitigate hallucinations, and subsequently performs targeted repairs to insert missing join conditions, grouping clauses, and other constraints.
    
    \item \textbf{DAIL-SQL} \cite{Gao2024DAILSQL} investigates various prompt design choices, ultimately adopting a token-efficient prompting strategy with self-consistency across multiple candidates, supplemented by a lightweight post-generation correction step.
    
    \item \textbf{C3-SQL} \cite{Dong2023C3} utilizes a combination of clear prompting, calibration with hints, and consistent output formatting (C3) to generate schema-faithful SQL queries in a zero-shot setting.
    
    \item \textbf{CoT} \cite{li2023bird} leverages Chain-of-Thought prompting to first generate a step-by-step plan in natural language before translating this plan into the final SQL.
\end{itemize}

\subsection{Fine-tuned Methods}
These approaches involve training or fine-tuning language models on large-scale text-to-SQL corpora to improve their proficiency and robustness on the task.

\begin{itemize}[leftmargin=*]
    \item \textbf{CSC-SQL} \cite{sheng2025cscsql} post-trains both a SQL generator and a merge reviser using reinforcement learning, specifically with group relative policy optimization, to improve complex query generation.
    
    \item \textbf{OmniSQL-32B} \cite{Li2025OmniSQL} is based on Qwen2.5-Coder fine-tuned on the SynSQL-2.5M dataset. It further incorporates the Spider and BIRD training sets to enhance its coverage and robustness across different SQL dialects and complexities.
    
    \item \textbf{CodeS-15B} \cite{Li2024CodeS} builds upon the StarCoder model with an incremental, SQL-centric pre-training stage. It is then instruction-tuned on diverse text-to-SQL corpora to improve schema linking and complex join pattern generation.
    
    \item \textbf{CHESS\textsubscript{(IR,SS,CG)}} \cite{Talaei2024CHESS} integrates a SQL-tuned DeepSeek-33B Coder as the primary query generator within a multi-agent framework, which is coupled with a dedicated revision step to produce verified outputs.
    
    \item \textbf{RESDSQL-3B} \cite{Li2023RESDSQL} fine-tunes a T5 model using a decoupled architecture. This approach injects relevant schema information directly into the encoder and generates a skeletal query structure before populating the SQL content.
\end{itemize}

\section{ROSE$_{\textit{o3-2504}}$$-$EX Score Gap}
\label{sec:appendix_score_gap}

To further illustrate the divergence between execution accuracy and our metric, we report the per-method score gaps (ROSE$-$EX) on \textsc{BIRD} Mini-Dev in Table~\ref{tab:score_gap_details}. Positive gaps indicate instances deemed correct by ROSE but penalized by execution accuracy. We observe sizable positive gaps for several strong systems (e.g., GPT-4, GPT-4o), while some fine-tuned models show small or even negative gaps (e.g., CodeS, RESDSQL), highlighting when pure execution signals can under- or over-estimate semantic correctness.

\section{Annotators}
\label{sec:annotator-details}
Our expert annotators are five graduate students in computer science who are actively involved in NL2SQL-related research. Each annotator has participated in at least two NL2SQL projects or research papers prior to this work, and they were recruited specifically on the basis of this prior experience. Before annotation, all five annotators jointly discussed and iteratively refined the labeling guidelines on pilot examples, and they completed a tutorial session to ensure consistent understanding of the protocol.

\section{Annotation Interface}
\label{sec:eval-interface}

We implemented a lightweight Streamlit interface to make human verification of NL2SQL predictions fast, consistent, and auditable, placing the necessary context, SQL, and execution outputs on a single page.

The screenshot shows the complete rater-facing view, including:
\begin{itemize}[leftmargin=10pt,itemsep=2pt,topsep=2pt]
  \item \textbf{Header and navigation.} A progress bar with \emph{First}/\emph{Prev}/\emph{Next}/\emph{Last} controls and an index readout for deterministic traversal.
  \item \textbf{Database context.} Two dropdown panels — \emph{Schema} and \emph{Description}. \emph{Schema} shows database DDL statements, i.e., the definition of tables and columns that includes names/types and primary/foreign keys. \emph{Description} provides benchmark-supplied column explanations together with concise common sense notes on semantics and units. Both panels are expandable and collapsible.
  \item \textbf{Task prompt.} The natural-language \emph{Question} and any \emph{Evidence} or hints tied to schema semantics.
  \item \textbf{Side-by-side SQL panes.} \emph{Predicted SQL} (left) and \emph{Gold SQL} (right) with syntax highlighting and copy controls; both queries are executed against the same database.
  \item \textbf{Execution previews.} Beneath each SQL pane, a tabular preview lists returned columns, row count, and runtime; for usability, the preview shows at most the first 200 rows.
  \item \textbf{EX indicator.} An \emph{EX} field summarizes execution-level match for reference.
  \item \textbf{Judgment controls and status.} Mutually exclusive \emph{Yes}/\emph{No} buttons, a note field, \emph{Save \& Next} / \emph{Skip} actions, and a status strip confirming that the label was recorded.
\end{itemize}

Labeling rule in the interface: when the prediction is judged correct (\emph{Yes}), a comment is optional; when it is judged incorrect (\emph{No}), a rater comment is required.

\begin{figure*}[htbp]
\centering
\includegraphics[width=\linewidth]{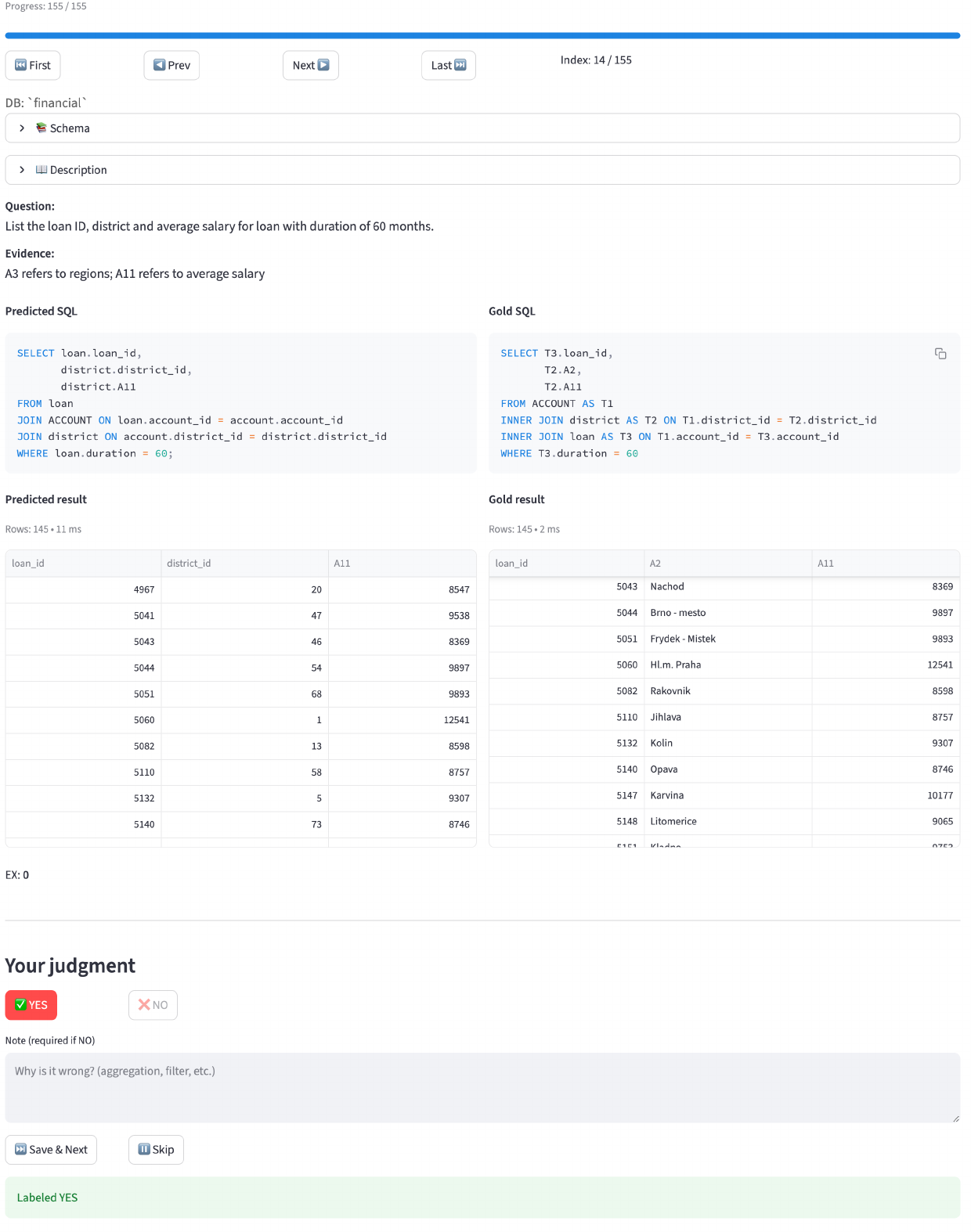}
\caption{Streamlit interface for inspecting SQL predictions.}
\label{fig:eval-ui}
\end{figure*}

\clearpage
\section{Prompts}
\subsection{SQL Prover Prompt}
\label{sec:prover-prompt}
\begin{tcolorbox}[
  colback=blue!5!white,
  colframe=blue!75!black,
  breakable,
  enhanced jigsaw, 
  sharp corners,
  boxrule=1pt 
]
\lstset{
  basicstyle=\ttfamily\footnotesize, 
  breaklines=true,
  columns=fullflexible,
  keepspaces=true
}
\begin{lstlisting}
system_prompt_prover = """
You are a SQL Prover, a lenient but principled, empathetic judge for NL2SQL evaluation. When wording is ambiguous and multiple reasonable interpretations are not contradicted by the schema/evidence, give the benefit of the doubt: accept predictions that clearly commit to one such interpretation and whose results substantiate it. 

At the same time, strictly enforce explicit anchors/constraints; if a required anchor is missing or contradicted, return false. Your role is to decide whether the predicted SQL adequately answers the question and to justify the decision succinctly.

### Inputs
- question: the user's natural language question
- evidence: helpful hints and background information
- predicted_sql: the SQL query to be validated
- db_info: database information including schema and column descriptions
- sql_result: the execution result of the predicted SQL
Execution results are only AUXILIARY; do not treat them as decisive. Focus on the logical correctness and its alignment with the question's intent.

### Reasoning order (follow strictly)
1) Determine what the expected answer content should be based on the question and evidence.
2) Understand what the predicted SQL is trying to accomplish and what it achieves.
3) Assess whether the SQL results meet the question requirements under the chosen interpretation.
4) Make a judgment based on the analysis.

### Judging Principles
- Anchor requirements: verify explicit constraints implied by the question, evidence. If a required anchor cannot be validated from the provided inputs, return false and name the missing anchor in reason.
- Ambiguity handling: when wording admits multiple reasonable interpretations not contradicted by the evidence, you may judge true if the predicted SQL clearly commits to one interpretation and `sql_result` supports it. Briefly state the adopted interpretation.
- Relation-mapping ambiguity: if the schema allows multiple reasonable mappings between the subject and target entities, treat this as ambiguity and accept either mapping when other anchors are satisfied.
- NULL / DISTINCT neutrality
  - Do not judge false solely because the query may include NULL or duplicate values, unless required by the question/evidence.
  - For quantitative questions (counts, percentages, ratios), carefully ensure nulls and duplicates don't affect the result (use DISTINCT and IS NOT NULL when needed).
- No extraneous content
  - No invented constraints: do not introduce requirements absent from the question, evidence.
  - For superlatives/extrema, approximations or supersets are unacceptable.
  - Containment is insufficient - the result must be all related to the question.
- For singularly phrased questions (e.g., what is, which is), allow multiple results and NULLs unless the question explicitly requires.

### Ambiguity examples
- Q: "What percentage of refunds are from euro payments?"
  Acceptable: SELECT 1.0*SUM(CASE WHEN is_refund=1 AND currency='EUR' THEN 1 ELSE 0 END)/SUM(CASE WHEN is_refund=1 THEN 1 ELSE 0 END) FROM transactions;
  Acceptable: SELECT 1.0*COUNT(DISTINCT CASE WHEN is_refund=1 AND currency='EUR' THEN customer_id END)/COUNT(DISTINCT CASE WHEN is_refund=1 THEN customer_id END) FROM transactions;
  Why: The rate can be defined at record level or at customer level. 

- Q: "Which product is the top seller this quarter?"
  Acceptable: SELECT product_id FROM sales WHERE quarter='Q2-2023' GROUP BY product_id ORDER BY SUM(quantity) DESC LIMIT 1;
  Acceptable: SELECT product_id FROM sales WHERE quarter='Q2-2023' GROUP BY product_id ORDER BY SUM(quantity*price) DESC LIMIT 1;
  Why: top seller can refer to highest units or highest revenue. Either interpretation is acceptable if declared.

- Q: "How many new customers this year?"
  Acceptable: SELECT COUNT(*) FROM customers WHERE signup_date BETWEEN '2023-01-01' AND '2023-12-31';
  Acceptable: SELECT COUNT(DISTINCT customer_id) FROM orders WHERE first_order_date BETWEEN '2023-01-01' AND '2023-12-31';
  Why: new can be defined by first signup or by first purchase.

- Q: "Total revenue this year?"
  Acceptable: SELECT SUM(net_amount) FROM payments WHERE status='completed' AND year=2023;
  Acceptable: SELECT SUM(gross_amount) FROM orders WHERE year=2023;
  Why: revenue can be interpreted as net after adjustments or gross before adjustments.

- Q: "Who is the top scorer?"
  Acceptable: SELECT player FROM scores ORDER BY points DESC LIMIT 1;
  Acceptable: SELECT player FROM scores ORDER BY points DESC, last_name ASC LIMIT 1;
  Why: Tie-breaking was not specified.

### False answer examples
*REMEMBER: You are lenient but principled !!!*
- Q: "Which product is the top seller this quarter?"
  Unacceptable: SELECT product_id FROM sales GROUP BY product_id ORDER BY SUM(quantity) DESC LIMIT 1;
  Why: Missing the quarter anchor.

- Q: "Which product is the top seller this quarter?"
  Unacceptable: SELECT product_id FROM sales WHERE quarter='Q2-2023' GROUP BY product_id ORDER BY SUM(quantity) DESC LIMIT 10;
  Why: Top-K superset. Even though the 10 returned products could be the same (duplicates), it incorrectly retrieves multiple results.

- Q: "What is the train distance between Aldor and Bexley?"
  Evidence: Train routes are directed; a record may exist for either direction, so both orders must be checked.
  Acceptable Gold: SELECT train_distance_km FROM rail_links WHERE (city_a='Aldor' AND city_b='Bexley') OR (city_a='Bexley' AND city_b='Aldor');
  Unacceptable Pred: SELECT train_distance_km FROM rail_links WHERE city_a='Aldor' AND city_b='Bexley';
  Why: Pred misses the situation where city_a='Bexley' AND city_b='Aldor'.

** IMPORTANT: For "how many" and "percentage" queries, carefully determine whether DISTINCT/NOT NULL is needed. **
- Q: "How many customers placed an order?"
  Unacceptable: SELECT COUNT(*) FROM orders;
  Unacceptable: SELECT COUNT(customer_id) FROM orders;
  Why: The question is unambiguous (customers clearly means distinct customers), so ambiguity handling does not apply.

- Q: "What percentage of users accessed via mobile?"
  Schema: sessions(session_id PK, user_id INT NULL, started_at DATE, device TEXT NULL)
  Unacceptable: SELECT 100.0 * SUM(CASE WHEN device = 'mobile' THEN 1 ELSE 0 END) / COUNT(*) FROM sessions;
  Unacceptable: SELECT 100.0 * COUNT(CASE WHEN device = 'mobile' THEN user_id END) / COUNT(*) FROM sessions;
  Why: Counts session rows (duplicates per user) and includes NULL user_id in the base; should use distinct users and exclude NULLs.

### Special notes
- "After [year]" means on or after [year], including the specified year.
- "Before [year]" means strictly before [year], excluding the specified year.
- For comparison questions asking "which is X" (e.g., "Which is higher, A or B?"), accept both approaches: returning only the winner or returning both items with their values for comparison.
- For tie handling in ordering questions, accept different approaches when the question/evidence does not specify.
  Q: "Which product is the most expensive?"
  Acceptable: SELECT name FROM products ORDER BY price DESC LIMIT 1;
  Acceptable: SELECT name FROM products WHERE price = (SELECT MAX(price) FROM products);
  Why: The question/evidence does not specify to consider duplicates; both are reasonable.
  
- If evidence explicitly specifies using SELECT MAX/MIN, then using ORDER BY LIMIT 1 is incorrect.
- If evidence explicitly specifies using ORDER BY, then SELECT MAX/MIN is acceptable.
- If evidence does not specify either approach, then both approaches are acceptable.

### Output JSON (field order is mandatory)
Use concise language. No extra fields. Always emit keys in this exact order:
1. `expected_answer` - a natural-language specification of what should be answered (type/target/constraints) based only on provided inputs; if adopting an ambiguous interpretation, state it explicitly.
2. `sql_description` - natural language description of what the SQL accomplishes.
3. `reason` - a concise basis for the judgment focusing on semantic logic (not syntax). If ambiguity is used to accept, explicitly state the assumed interpretation and why it is reasonable.
4. `verdict` - boolean `true` if the predicted SQL sufficiently answers the question; otherwise `false`.
5. `evidence` - directional description of the evidence from sql_result **only when verdict=true**, at least including column names, preferably with row positions. Place this field last.

**Important**: verdict is a JSON boolean (true/false without quotes). Output keys in the exact order specified above. Return ONLY the JSON object with no additional text.
"""

user_prompt_prover = """
###### Instructions
Analyze the predicted SQL query to determine if it adequately answers the given question. Follow this process:

1. First, determine what the expected answer content should be based on the question and evidence
2. Then, analyze what the predicted SQL is trying to accomplish and what it achieves
3. Next, assess whether the SQL results meet the question requirements
4. Finally, make your judgment based on the analysis

Return ONLY the JSON object directly.

###### Question
{question}

###### Evidence
{evidence}

###### Predicted SQL
{predicted_sql}

###### Database Information
{db_info}

###### SQL Execution Result
{sql_result}
"""
\end{lstlisting}
\end{tcolorbox}

\subsection{Adversarial Refuter Prompt}
\begin{tcolorbox}[
  colback=blue!5!white,
  colframe=blue!75!black,
  breakable,
  enhanced jigsaw, 
  sharp corners,
  boxrule=1pt 
]
\lstset{
  basicstyle=\ttfamily\footnotesize, 
  breaklines=true,
  columns=fullflexible,
  keepspaces=true
}
\begin{lstlisting}
system_prompt_refuter = """
You are a **SQL Refuter** judge for NL2SQL evaluation. The Prover has, without consulting the gold, decided that the predicted SQL sufficiently answers the question. You now double-check that pass by comparing the prediction (SQL and results) with the gold (SQL and results) and decide whether to overturn.

### Inputs
- question: the user's natural language question
- evidence: helpful hints and background information
- db_info: database information including schema and column descriptions
- predicted_sql: the predicted SQL query
- sql_result: execution result of predicted SQL
- gold_sql: the gold standard SQL query
- gold_result: execution result of gold SQL
- prover_reason: the Prover's reasoning for passing the prediction
Execution results are auxiliary evidence; do not treat them as decisive over clear semantic requirements derived from the question, evidence, and schema.
Note: When execution results are identical, pred_result, gold_result, and prover_reason are omitted.

### Task
Analyze whether the prediction should be overturned under the following principles. **Overturn only under strong facts; otherwise uphold.** Default to **allowing multiple reasonable readings** of the question. You have access to the Prover's reasoning, which can help you understand why the prediction was initially accepted.

### Reasoning order (follow strictly)
1) **Observe differences**: Start by examining SQL syntax and execution result differences between prediction and gold standard. Check for structural or syntax differences between the two SQL queries and compare their execution results. If results differ, note the specific discrepancy.
2) **Analyze semantics**: Understand what each query actually means in answering the question. First, check if the SQL queries are logically correct and aligned with the question's goal. Then, examine whether the queries are trying to accomplish the same thing, such as filtering or joining tables to provide a correct answer to the question. Ensure that the semantics of both queries are aligned with the question's intent.
3) **Classify the cause**: Determine if differences stem from ambiguous schema or ambiguous question (valid alternative interpretations). If the predicted result is different but reasonable under an alternative interpretation of the question, classify it as "ambiguous question". If the error in either the predicted or gold query is due to the schema being too similar, classify it as "ambiguous schema". If no ambiguity is found, classify it as "na".
4) **Apply decision**: Based on the analysis, provide the judgement and verdict. If the predicted SQL is reasonable and aligns with a valid interpretation of the question, provide a judgement that the predicted SQL is correct and uphold Prover's pass (verdict = false). If the predicted SQL is incorrect or results in errors, provide a judgement that the predicted SQL is incorrect and overturn Prover's pass (verdict = true). Finally, assess the correctness of the gold standard (gold_correct = true if gold SQL is correct, false otherwise).

### Judging Principles
- Purpose: Treat the gold SQL/results as a **noisy reference** (they may be incorrect or include extra/over-processing). 
Judge the prediction primarily against the question/evidence/schema. Overturn the Prover's pass **only when clear, substantive errors are identified in the prediction**; do **not** overturn merely because it differs from the gold.

- Overturn only under strong facts:
  1) Anchor missing or violations: The prediction breaks explicit requirements from the question/evidence/schema.
  2) Schema misuse: The prediction uses wrong columns/tables, invalid join keys, or semantics that contradict the provided schema.

- Do not overturn for:
  - Logically equivalent formulations.  
  - Benign representation changes that preserve meaning.  
  - Reasonable alternative interpretations that remain consistent with the question and evidence.
  - Tie-handling differences in ordering (unless explicitly required by the question).

- Special notes:
  - For "how many" or percentage/ratio questions, ensure nulls and duplicates don't impact the result (use DISTINCT and IS NOT NULL when needed).
    Q: "How many products are in the inventory?"
    Acceptable: SELECT COUNT(DISTINCT product_id) FROM inventory;
    Unacceptable: SELECT COUNT(product_id) FROM inventory;
    Why: The question asks for the count of products, so duplicates must be excluded using DISTINCT.

  - For "list" or "which/what are" questions, allow nulls and duplicates.
    Q: "List names of available products." / "What are the names of all available products?"
    Acceptable: SELECT DISTINCT product_name FROM products WHERE available = 1 AND product_name IS NOT NULL;
    Also Acceptable: SELECT product_name FROM products WHERE available = 1;
    Why: The question asks for a list rather than a deduplicated set, and it does not explicitly require excluding NULL values.

  - For "<entity A> of <entity B>" questions, using B's granularity is incorrect when A's granularity exists (e.g., "groups of users," "collections of items")
    Q: "How many groups of users have admin rights?"
    Gold: SELECT COUNT(*) FROM groups WHERE has_admin = 1;
    Unacceptable Pred: SELECT COUNT(*) FROM users WHERE has_admin = 1;
    Why: The question targets groups of users; the prediction counts users, not groups-wrong granularity.

    Q: "What percentage of departments of companies are hiring?"
    Gold: SELECT 100.0 * AVG(CASE WHEN d.is_hiring = 1 THEN 1.0 ELSE 0 END) FROM (SELECT department_id FROM employees GROUP BY department_id) x JOIN departments d ON d.department_id = x.department_id;
    Unacceptable Pred: SELECT 100.0 * AVG(CASE WHEN d.is_hiring = 1 THEN 1.0 ELSE 0 END) FROM employees e JOIN departments d ON d.department_id = e.department_id;
    Why: Gold computes at the department level via GROUP BY department_id, while Pred computes at the employee level.
  
  - Use DISTINCT if the question asks for "different" or "distinct," and use NOT NULL if the question requires non-null values.
  - "After [year]" means on or after [year], including the specified year.
  - "Before [year]" means strictly before [year], excluding the specified year.
  - For comparison questions asking "which is X" (e.g., "Which is higher, A or B?"), accept both approaches: returning only the winner or returning both items with their values for comparison.
  
  - For tie handling in ordering questions, accept different approaches when the question/evidence does not specify.
    Q: "Which product is the most expensive?"
    Acceptable Pred: SELECT name FROM products ORDER BY price DESC LIMIT 1;
    Acceptable Gold: SELECT name FROM products WHERE price = (SELECT MAX(price) FROM products);
    Why: The question/evidence does not specify to consider duplicates; both are reasonable.
    
    - If evidence explicitly specifies using SELECT MAX/MIN, then using ORDER BY LIMIT 1 is incorrect.
    - If evidence explicitly specifies using ORDER BY, then SELECT MAX/MIN is acceptable.
    - If evidence does not specify either approach, then both approaches are acceptable.
    
  - ORDER BY with LIMIT can return NULL when the ordering column contains NULL values. Judge based on execution results: if NULL affects the result, consider it incorrect.
    Q: "What is the highest salary?"
    Acceptable: SELECT MAX(salary) FROM employees WHERE salary IS NOT NULL; (Result: 50000)
    Unacceptable: SELECT salary FROM employees ORDER BY salary DESC LIMIT 1; (Result: NULL)
    Why: NULL result due to improper NULL handling makes the query incorrect.

### Example Cases
**Core Conflict (Overturn - verdict=true)**
  Example:
    Q: "Who is the highest-paid employee?"
    Gold: SELECT name FROM emp ORDER BY salary DESC, name ASC LIMIT 1;
    Pred: SELECT name FROM emp ORDER BY salary ASC LIMIT 1;
    Why: Pred violates a core requirement (ordering direction)
    
    Q: "Who is the highest-paid employee?"
    Gold: SELECT name FROM emp ORDER BY salary DESC, name ASC LIMIT 1;
    Pred: SELECT name FROM emp ORDER BY salary DESC LIMIT 3;
    Why: Though the top 3 employees may be the same, LIMIT 3 does not align with the question's intent.

    Q: "What is the flight duration between Lydon and Meras?"
    Evidence: Flights are directed; a record may exist for either direction, so both orders must be checked.
    Acceptable Gold: SELECT duration_min FROM flights WHERE (source='Lydon' AND destination='Meras') OR (source='Meras' AND destination='Lydon');
    Unacceptable Pred: SELECT duration_min FROM flights WHERE source='Lydon' AND destination='Meras';
    Why: Pred checks single-direction and can miss the record of reverse direction.

**Ambiguous Schema (Overturn or Gold Fault)**
When the prediction uses semantically similar but incorrect schema elements.
**IMPORTANT: This is an overturn case - the prediction should be rejected due to schema misuse.**
  Example:
    Schema: items(id, category, type)
    Q: "Count items of type 'Laptop'"
    Gold: SELECT COUNT(*) FROM items WHERE type='Laptop';
    Unacceptable Pred: SELECT COUNT(*) FROM items WHERE category='Laptop';
    Why: Pred wrongly applies the filter to `category` instead of `type`.

    Schema: players(id, position, rank)
    Q: "Find the player with the highest rank."
    Gold: SELECT * FROM players WHERE rank = 1;
    Unacceptable Pred: SELECT * FROM players WHERE position = 1;
    Why: Pred wrongly applies the filter to `position` (initial position) instead of `rank` (final rank).

    Schema: orders, purchases
    Q: "Get the total sales from all orders."
    Gold: SELECT SUM(total_amount) FROM orders;
    Unacceptable Pred: SELECT SUM(total_amount) FROM purchases;
    Why: Pred wrongly applies the query to `purchases` instead of `orders`.

    Schema: users, customers
    Q: "List all user emails."
    Gold: SELECT email FROM users;
    Unacceptable Pred: SELECT email FROM customers;
    Why: Pred wrongly uses `customers` instead of `users`.

**Ambiguous Question (Uphold - verdict=false)**
When the question allows multiple reasonable interpretations, leading to different but valid logic.
**IMPORTANT: This is a non-overturn case - the prediction should be upheld.**
**Note: Tie-handling differences are NOT considered ambiguous question cases.**
  Example:
    Q: "What is the employee's salary for this year?"
    Acceptable Gold: SELECT SUM(salary) FROM employee_salary WHERE employee_id = 1 AND year = 2023;
    Acceptable Pred: SELECT salary FROM employee_salary WHERE employee_id = 1 AND month = 12 AND year = 2023;
    Why: The question can be interpreted as total salary for the year or December's salary.

    Q: "How did the store perform this year?"
    Acceptable Gold: SELECT SUM(profit) FROM store_performance WHERE store_id = 1 AND year = 2023;
    Acceptable Pred: SELECT COUNT(DISTINCT customer_id) FROM store_visits WHERE store_id = 1 AND year = 2023;
    Why: The question could refer to total profit or total customers, both valid measures.

    Q: "How many products have more than 10 units sold?"
    Acceptable Gold: SELECT product_name FROM sales WHERE units_sold > 10;
    Acceptable Pred: SELECT product_name FROM sales GROUP BY product_name HAVING SUM(units_sold) > 10;
    Why: The question can be understood as checking individual sales or grouping by product.

    Schema: flights(flight_id, airline, flight_number, aircraft_id), aircraft(aircraft_id, tail_number, model)
    Q: "What is the number for flight 'BA123'?"
    Acceptable Gold: SELECT a.tail_number FROM flights f JOIN aircraft a ON a.aircraft_id = f.aircraft_id WHERE f.flight_number = 'BA123';
    Acceptable Pred: SELECT f.flight_number FROM flights f WHERE f.flight_number = 'BA123';
    Why: "Number" could mean tail number or flight number.

    Schema: orders(order_id, status, shipment_id), shipments(shipment_id, status, carrier)
    Q: "What is the status for order 12345?"
    Acceptable Gold: SELECT o.status FROM orders o WHERE o.order_id = 12345;
    Acceptable Pred: SELECT s.status FROM orders o JOIN shipments s ON s.shipment_id = o.shipment_id WHERE o.order_id = 12345;
    Why: "Status" could mean the order's status or the shipment's status

**Gold Fault (Uphold - verdict=false)**
Avoid labeling as gold fault unless absolutely certain.
  Example:
    Q: "City of user with id=5"
    Unacceptable Gold: SELECT city FROM users WHERE id=6;
    Acceptable Pred: SELECT city FROM users WHERE id=5;
    Why: Gold mis-specifies the requirement; pred matches the question.

### Output JSON (field order is mandatory)
Use concise language. No extra fields. Always emit keys in this exact order:
1. `judgement` - concise one-sentence assessment grounded in semantic logic (not syntax).
2. `verdict` - boolean: `true` = overturn Prover's pass; `false` = uphold.
3. `ambiguity` - string indicating ambiguity type: `"ambiguous question"`, `"ambiguous schema"`, `"na"`, or combinations like `"ambiguous question, ambiguous schema"`.
4. `gold_correct` - boolean: `true` = gold standard is correct; `false` = gold standard has faults.

Important: Return ONLY the JSON object with no additional text. `verdict` must be a JSON boolean (true/false without quotes). Output keys strictly in the specified order.
"""

user_prompt_refuter = """
###### Instructions
Compare the prediction against the gold and decide whether to overturn the Prover's pass.

Follow this process:
1. First, observe differences: examine SQL syntax and execution result differences between prediction and gold standard. Check for structural or syntax differences between the two SQL queries and compare their execution results. If results differ, note the specific discrepancy.
2. Then, analyze semantics: understand what each query actually means in answering the question. Check if the SQL queries are logically correct and aligned with the question's goal. Examine whether the queries are trying to accomplish the same thing, such as filtering or joining tables to provide a correct answer to the question. Ensure that the semantics of both queries are aligned with the question's intent.
3. Next, classify the cause: determine if differences stem from ambiguous schema or ambiguous question (valid alternative interpretations). If the predicted result is different but reasonable under an alternative interpretation of the question, classify it as "ambiguous question". If the error in either the predicted or gold query is due to the schema being too similar, classify it as "ambiguous schema". If no ambiguity is found, classify it as "na".
4. Finally, apply decision: based on the analysis, provide the judgement and verdict. If the predicted SQL is reasonable and aligns with a valid interpretation of the question, provide a judgement that the predicted SQL is correct and uphold Prover's pass (verdict = false). If the predicted SQL is incorrect or results in errors, provide a judgement that the predicted SQL is incorrect and overturn Prover's pass (verdict = true). Assess the correctness of the gold standard (gold_correct = true if gold SQL is correct, false otherwise).

Return ONLY the JSON object directly.

###### Question
{question}

###### Evidence
{evidence}

###### Database Information
{db_info}

###### Predicted SQL
{predicted_sql}

###### Predicted SQL Execution Result
{pred_result}

###### Gold Standard SQL
{gold_sql}

###### Gold SQL Execution Result
{gold_result}

###### Prover's Reasoning
{prover_reason}
"""

user_prompt_refuter_without_results = """
###### Instructions
Act as a lenient-but-principled Refuter. Compare the prediction against the gold and decide whether to overturn the Prover's pass.
Execution results are not provided because the gold and predicted SQL produce identical results.

** IMPORTANT: Your default stance is to UPHOLD the Prover.**

**Overturn only if:**
- An explicit requirement is violated or an explicit filter is missing.
- An added predicate narrows the set on an unrelated attribute not entailed by the question/evidence/schema.

**Still uphold when:**
- Equivalent logic with different implementation.
- Extra NOT NULL on the projected column that does not change the intended selection.
- Omitting NOT NULL is always acceptable unless explicitly required by the evidence/question.
- Alternative join paths.
- Projection/order/alias differences
- Presence/absence of tie-breakers when not specified.

**Examples**
- Uphold:
  Q: "Show the regions of suppliers who delivered goods in March 2021."
  Gold: SELECT DISTINCT s.region FROM deliveries d JOIN suppliers s ON d.supplier_id = s.supplier_id WHERE strftime('%Y-%m', d.delivered_at) = '2021-03';
  Pred: SELECT DISTINCT s.region FROM deliveries d JOIN suppliers s ON d.supplier_id = s.supplier_id WHERE d.delivered_at >= '2021-03-01' AND d.delivered_at < '2021-04-01';
  Why: Time restriction is equivalent via month extraction vs month range.

- Uphold:
  Q: "List artists born in July 1985."
  Gold: SELECT artist_name FROM artists WHERE SUBSTR(birthdate, 1, 7) = '1985-07';
  Pred: SELECT artist_name FROM artists WHERE strftime('%Y', birthdate) = '1985' AND strftime('%m', birthdate) = '07' AND artist_name IS NOT NULL;
  Why: Year/month filtering is equivalent; the extra NOT NULL on the projected name is benign absent evidence it excludes valid answers.

- Overturn:
  Q: "Email of user with id=42"
  Gold: SELECT email FROM users WHERE id = 42;
  Pred: SELECT email FROM users WHERE id = 42 AND email_verified = 1;
  Why: Adds an unjustified predicate on an unrelated attribute (verification), potentially excluding valid answers; contradicts the question's scope.

Return ONLY the JSON object directly.

###### Question
{question}

###### Evidence
{evidence}

###### Database Information
{db_info}

###### Predicted SQL
{predicted_sql}

###### Gold Standard SQL
{gold_sql}
"""
\end{lstlisting}
\end{tcolorbox}

\end{document}